 \newlength\smallfigwidth
 \newlength\tinyfigwidth
 \documentclass[aps,prb,twocolumn,floatfix,amsmath,amssymb,showpacs,showkeys]{revtex4}
\newcommand{\be}{\begin{equation}}
\newcommand{\ee}{\end{equation}}
\newcommand{\bn}{\begin{eqnarray}}
\newcommand{\en}{\end{eqnarray}}

\newcommand{\sgn}{\, {\rm sgn} }
\newcommand{\ew}{$\epsilon(\omega)$\ }

\usepackage{graphicx}
\usepackage[hypertex]{hyperref}
\begin{document}

\preprint{KSU-Wysin}

\title{Effects of Interband Transitions on Faraday Rotation in Metallic Nanoparticles}
\author{G.\ M.\  Wysin}
\email{wysin@phys.ksu.edu}
\homepage{http://www.phys.ksu.edu/personal/wysin}
\affiliation{Kansas State University, Manhattan, KS 66506-2601}
\author{Viktor Chikan}
\email{chikan@ksu.edu}
\affiliation{Kansas State University, Manhattan, KS 66506-2601}
\author{Nathan Young}
\email{youngnt@rose-hulman.edu}
\altaffiliation[Permanent address: ]{Rose-Hulman Institute of Technology, Terre Haute, IN 47803.}
\affiliation{Kansas State University, Manhattan, KS 66506-2601}
\author{Raj Kumar Dani}
\email{rdani@ksu.edu}
\affiliation{Kansas State University, Manhattan, KS 66506-2601}
\date{\today}
\begin{abstract}
{The Faraday rotation in metallic nanoparticles is considered based on a quantum
model for the dielectric function \ew in the presence of a DC magnetic field $B$. 
We focus on effects in \ew due to interband transitions (IBTs), which are
important in the blue  and ultraviolet for noble metals used in plasmonics.
The dielectric function is found using the perturbation of the electron density matrix
due to the optical field of incident electromagnetic radiation.  The calculation is
applied to transitions between two bands ($d$ and $p$, for example) separated by a gap, as 
one finds in gold  at the L-point of the Fermi surface.  The result of the 
DC magnetic field is a shift in the effective optical frequency causing IBTs
by $\pm \mu_B B / \hbar$, where opposite signs are associated with left/right 
circular polarizations.  Faraday rotation for a dilute solution of 17 nm diameter gold 
nanoparticles is measured and compared with both the IBT theory and a simpler Drude model 
for the bound electron response.  Effects of the plasmon resonance mode on Faraday rotation 
in nanoparticles are also discussed.}
\end{abstract}

\maketitle

\section{Introduction: Faraday rotation enhancement and plasmon modes}
There is great interest to design new materials with enhanced Faraday rotation\cite{Hecht97} (FR); 
these media\cite{Kim03,Bam04,Hay08,Yu08} are good candidates for applications like field detectors, 
phase modulators and optical isolators.   Nanoparticles (NPs) of radius $a$ much less than the 
wavelengths $\lambda$ of the electromagnetic (EM) radiation are particularly interesting 
possibilities,\cite{Kelly+03} because one hopes to be able to tune their fundamental physical properties
that determine the dielectric permittivity $\epsilon(\omega)$, which influences the FR signal. Further, 
metallic NPs or NPs with metallic shell coatings have a surface plasmonic mode\cite{Mul96} where the electron 
response is greatly enhanced, which leads to increased FR response.\cite{Dani11}  In magnetic core 
NPs with metallic shell coatings the surface plasmonic mode interacts with a transition in the 
magnetic core, which is another process that leads to enhanced FR.\cite{Jain++09} 
Faraday rotation and circular dichroism in NP aggregates of various geometries\cite{Pakdel+12} offers
promise for control of magneto-optical effects.  Backscattering of light in disordered media
may lead to enhanced Faraday rotation effects.\cite{Gas+13}

It is simplest to use a classical Drude term to represent approximately the dielectric response of 
bound electrons, \cite{Dani11} or even ignore the bound electron dynamic response,\cite{GuKornev10}
however, these approaches do not describe the dielectric properties well at higher frequencies.
Hui and Stroud\cite{Hui+87} have considered the FR response of a dilute suspension of small particles, with a 
Drude approximation for NP dielectric function.
It is the goal here to compare a classical phenomenological Drude approach with a quantum
model appropriate for noble metals such as gold,\cite{Inouye++98,Scaffardi06} where interband transitions 
(IBTs) take place from $d$ to $sp$ bands at the L-point.\cite{CS71}  We consider a case where the 
NPs are sufficiently small, so that the primary effect of the DC magnetic field that produces 
Faraday rotation is a Zeeman splitting of the band states, rather than an entire series of 
Landau levels.\cite{Arun69} 

The surface plasmon frequency $\omega_{sp}$ in a 
NP in the Rayleigh limit ($a\ll \lambda$) is considerably less than the bulk plasmon 
frequency $\omega_p$ for the same metal.  This is due partly to a geometrical effect, but the
more significant reason is that interband transitions taking place above a 
gap energy greatly modify the dielectric function in the region of the plasmon resonance.
For gold, the bulk plasmon, well into the ultraviolet at 138 nm, is moved 
to around 520--532 nm for the surface plasmons of NPs.\cite{Dani11} But to get 
the correct description, the interband transitions must be taken into account for 
describing $\epsilon(\omega)$.  Here, we include IBTs for the bound electron contribution 
to $\epsilon(\omega)$, in the presence of a DC magnetic field, so that the Faraday 
rotation properties can be described.

Faraday rotation is a magneto-optical phenomenon\cite{Hecht97} that measures the fundamental 
electronic, optical and magnetic response of a dielectric medium. It is similar to optical 
rotation\cite{Jones61} except that FR requires an applied magnetic field.  The Faraday 
rotation is the change in the polarization of an EM wave as it propagates through some 
medium parallel to the axis of a quasi-static magnetic field.  The interaction between the 
DC field and the charges in the medium leads to different speeds of propagation and 
different wave vectors $k_R$ and $k_L$ for the right and left circular polarization components 
of EM waves,  
leading to the net rotation of initially linearly polarized waves.  

The basic parameter to describe the degree of FR is the Verdet factor, 
$\upsilon$, which is the rotation angle of the polarization per
unit propagation length $z$ per applied magnetic field $B$:
\be
\label{ups}
\upsilon = \frac{\varphi}{Bz} \  .
\ee
We consider a composite medium of gold NPs in water.  At small volume fraction of gold, $f_s \ll 1$,
the Verdet factor is linear in the volume fraction.  Generally, the rotation $\varphi$ is also 
linear in magnetic field at low enough fields (although materials of other symmetry\cite{KBE85} 
can exhibit quadratic dependence on $B$). Then $\upsilon$ does not depend on $B$, and 
we consider only this regime.  For dilute composites, the Verdet factor per volume fraction is 
a better quantity for consideration, defined as
\be
\Upsilon \equiv \frac{\upsilon}{f_s} = \frac{\varphi}{B z f_s} \ .
\ee

The Faraday rotation results from the phase difference of the polarizations, 
\be
\label{fr}
\varphi = \tfrac{1}{2} {\rm Re}\left\{ k_R-k_L\right\} z \ ,
\ee
where the respective wave vectors for propagation of the two circular 
polarizations are determined by relative dielectric functions 
$\epsilon_R$ and $\epsilon_L$,
\be
\label{kvec}
k_{R,L} = \frac{\omega}{c} \sqrt{\mu\epsilon_{R,L}} \ .
\ee
with $\mu$ being the relative magnetic permeability of the medium and $c$ being the 
speed of light in vacuum.
The theoretical description of the FR signal in some metallic or metallic-shell NPs, 
as a result, requires an accurate description of the dielectric function \ew for
the metal, including the presence of the DC magnetic field.  Especially, it is important
to have a reliable description of \ew at shorter wavelengths, into the ultraviolet,
below the plasmon frequency for the NPs.  Indeed, in order for the theory to correctly
predict the plasmon frequency $\omega_{sp}$, requires knowing \ew all the way into
the ultraviolet.  Thus, the goal here is to get an accurate theory for \ew with
the DC magnetic field present, that includes interband transitions for the 
bound electrons as well as the usual plasmon response of the free electron gas.

Below we begin by describing the NP synthesis and a description of the experimental
measurement of the Faraday rotation.
Then we will continue by summarizing  the basic relations among the dielectric functions 
$\epsilon(\omega)$, $\epsilon_L(\omega)$, and $\epsilon_R(\omega)$ and the FR response.  
(All $\epsilon$ are understood to be relative dielectric functions, leading to
index of refraction,  $n=\sqrt{\mu\epsilon}$.) Next, a simple classical model for 
\ew using bound electrons and based on the Drude model is described briefly, 
for comparison with the quantum calculation for the IBTs.  In that model the effect 
of IBTs is approximated by the response function of a set of bound electrons, 
with some binding frequency $\omega_0$.  Part of the motivation for the quantum 
calculation of the IBTs is to determine the validity of this simpler classical model. 

For the quantum effects of IBTs, we adopt the approach used by Boswarva \textit{et al.}\cite{BHL62} 
and also by Adler\cite{Adler62} of finding the perturbations of the electron density matrix that are
caused by only the electric field of the EM waves.  The optical magnetic field is ignored. However, 
the DC magnetic field enters because it shifts the band states.  This is a simple Zeeman shift; 
for nanometer-sized systems there is no sense to Landau levels that were used by Boswarva 
\textit{et al.}, and also in theory developed by Halpern \textit{et al.},\cite{Halpern+64}
due to the geometric confinement (the NP radii are smaller than the Landau
radius $r_0=\sqrt{\tfrac{2\hbar}{eB}}$, for applied magnetic induction $B$).  
Following through the calculation, there is a contribution to \ew that requires summing over 
IBTs with a range of energies.  Those integrals are evaluated in two different models: 
A three-dimensional (3D) band model and a 1D band model.  The results are presented both  
with the presence of a phenomenological electron damping constant $\gamma$ and in a limit
that this damping goes to zero.  The results are also compared with earlier calculations of
the IBT contribution by Inouye {\it et al.}\cite{Inouye++98} and by Scaffardi and
Tocho,\cite{Scaffardi06} that did not include the DC magnetic field.

The net \ew includes contributions from both the bound (IBTs) and free (electron gas)
electrons. We apply the results to calculate the scaled Verdet factor $\Upsilon$ 
for a dilute solution of solid metallic NPs, using the parameters for gold in water.  
The effects of the dilution are considered most simply by the Maxwell-Garnett theory,
\cite{MG04,BW65} assuming that the NPs do not aggregate.
We find that even with the quantum IBTs included for bound electrons, the experimentally
measured FR signal is about $10 \times$ stronger than that predicted by the theory.

We will conclude with comments on the applicability of the results for other systems
with plasmonic enhancements of dielectric responses.

\section{Synthesis of gold NPs and Faraday rotation measurements}
\label{synthesis}
The large gold nanoparticles are prepared 
from the reduction of HAuCl$_4$ solution by sodium citrate solution as described by 
Turkevich {\it et al}.\cite{Turk+51} Briefly, 5 mg of HAuCl$_4$ and 50 mg of 
sodium citrate are dissolved in 95 ml and 5 ml of doubly distilled water, respectively. 
The HAuCl$_4$ solution is heated to about 70$^{\circ}$ C and the sodium citrate solution is
added, vigorously stirring the solution for 50 minutes. The color of the solution gradually
changes from faint pink to wine red. The resulting large gold nanoparticles have size 
$17 \pm 3$ nm.  Assuming 100\% reduction of the gold into NPs, the upper limit of volume 
fraction of gold in the solution is $f_s=1.50\times 10^{-6}$.  By analyzing the extinction 
coefficient of the solution by the techniques in Ref.\ \onlinecite{Liu+07}, the actual 
volume fraction is estimated to be $f_s=1.23\times 10^{-6}$. 

The Faraday rotation spectrum of NPs in water solution 
was measured with the help of a home built pulsed magnet. The magnet consists of
a helical coil machined from a copper beryllium block and electroplated with silver. The pulsed
current to the coil is provided via a simple RLC circuit. The capacitor bank of 77.3 $\mu$F from
Maxwell Laboratories is charged by a power supply/charger of Lumina Power, Inc. The
power supply uses 100-240 V AC-50/60 Hz input and output of 10 kV at 500 J/s in continuous
operation. The charge from the capacitor bank is discharged into the coil via a high voltage
trigger spark gap. The current is monitored in the circuit via a Rogowski coil, which measures
the current derivative. 

The Faraday rotation of the nanoparticle solutions is measured in a plastic
cell placed in the coil. A flash light source is triggered along with the pulsed magnet that allows
the synchronization of the magnet with the optical measurement. The duration of the light pulse
is 1.4 $\mu$s, while the duration of the magnetic pulse is $\sim 50~\mu$s, which allows that 
during the optical measurement the magnetic field is relatively constant. In front of the flash 
light source a polarizer is placed to produce polarized light for the Faraday measurement. 
The polarized light passes through the sample containing the nanoparticle solutions. The light 
leaving the optical cell passes through another polarizer that is set to 45 degree with respect 
the first polarizer. The light then enters a fiber optic spectrometer, which is also synchronized 
with the pulsed magnet and the light source. The Faraday rotation is calculated from the intensity 
change in the spectrum before and after the magnetic pulse. The magnetic field and the Faraday 
rotation setup are calibrated with water placed into the optical cell. The measurements are taken 
at 4.2 tesla magnetic fields.

\section{Theory: Dielectric polarization, currents, \ew  and  Faraday rotation}
We consider EM radiation at frequency $\omega$ with the electric field 
${\bf E}(t)\sim e^{-i\omega t}$, incident on a  material particle (an individual 
NP) much smaller then the wavelength (the Rayleigh limit).  Then the field 
${\bf E}$ is taken as uniform inside the sample.  The dielectric properties 
are based on the averaged dipole moment of the electrons of charge $e$, 
${\bf d}=e{\bf r}$. For $n=N/V$ electrons per unit volume, the electric 
polarization can be expressed as
\be
{\bf P} = n\langle{\bf d}\rangle  = \tilde\chi \cdot \epsilon_0 {\bf E} \ ,
\ee
where $\epsilon_0$ is the permittivity of vacuum and $\tilde\chi$ is the
susceptibility tensor that is to be found.  The dielectric function considered
as a tensor $\tilde\epsilon$ is defined via the electric displacement 
${\bf D}=\epsilon_0 \tilde{\epsilon} \cdot {\bf E}$ or
\be
{\bf D}=\epsilon_0 {\bf E}+{\bf P} \ ,
\ee
from which the usual definition results,
\be
\tilde{\epsilon} = {\bf 1}+\tilde\chi \ .
\ee
It is useful to realize another way to get to $\tilde\epsilon$,
via averaging of the microscopic currents, i.e., those caused by the 
optical fields.  The dielectric medium under study has current density 
${\bf J}$, which combines with the vacuum  displacement current. 
In this view the Ampere/Maxwell Law is  
\be
{\bf \nabla \times  H} = {\bf J} + \epsilon_0 \frac{\partial \bf E}{\partial t} \ .
\ee
All the effects of the medium are contained in ${\bf J}$.  This must
be equivalent to the alternative viewpoint that the currents are represented
instead by a dielectric function, 
\be
{\bf \nabla \times  H} = \frac{\partial \bf D}{\partial t} \ .
\ee
Considered at the frequency of the EM radiation with time derivatives 
$\partial/\partial t \rightarrow -i\omega$, these alternate views give
\be
{\bf J}  = -i\omega \epsilon_0 \left( \tilde\epsilon -{\bf 1}\right)  \cdot {\bf E}
= -i\omega \tilde\chi \cdot \epsilon_0{\bf E} \ .
\ee
Thus, an averaging of the microscopic currents will also lead to the susceptibility
and dielectric tensors.

We assume that the DC magnetic field ${\bf B}$ is along the $\hat{z}$-direction,
the same as the propagation direction of the EM waves, with wave vector ${\bf k}=k\hat{z}$.  
Then the electric field in the waves has only $xy$ components; only the transverse 
part of the dielectric tensor is needed.  In this situation it has the 
following symmetry\cite{Xia90}
\be
\label{eps-sym}
\tilde\epsilon =  \left[ \begin{array}{cc} \epsilon_{xx} & \epsilon_{xy}
\\ -\epsilon_{xy} & \epsilon_{xx} \end{array} \right]
=\left[ \begin{array}{cc} \epsilon_{xx} & i {\cal E}_{xy}
\\ -i {\cal E}_{xy} & \epsilon_{xx} \end{array} \right] \ .
\ee
The off-diagonal elements are determined by the DC magnetic field; they vanish when
${\bf B}=0$. The variable ${\cal E}_{xy}=-i\epsilon_{xy}$ is  convenient later; 
it is real in the absence of electron damping. 
The EM waves that propagate without any change in polarization are those with 
polarization vectors that are eigenvectors of $\tilde\epsilon$.  These eigenstates
are the usual states of right and left circular polarization.  Thus, solving
the eigenvector problem, 
$\tilde\epsilon \cdot \hat{\bf u}_i = \epsilon_i \hat{\bf u}_i,\ i=1,2$, 
with eigenvalues $\epsilon_i$ and eigenvectors $\hat{\bf u}_i$,
one finds the right circular polarization state (negative helicity) with $E_y=-iE_x$:
\be
\label{epsR}
\epsilon_R = \epsilon_{xx}+{\cal E}_{xy}\ , \quad 
\hat{\bf u}_R=\tfrac{1}{\sqrt{2}}(\hat{x}-i\hat{y}) \ ,
\ee
and the left circular polarization state (positive helicity) with $E_y=+iE_x$:
\be
\label{epsL}
\epsilon_L = \epsilon_{xx}-{\cal E}_{xy}\ , \quad 
\hat{\bf u}_L=\tfrac{1}{\sqrt{2}}(\hat{x}+i\hat{y}) \ .
\ee
Each mode has a different wave vector for propagation, according to expression 
(\ref{kvec}).  Then starting from a linearly polarized wave at position $z=0$, 
its right and left circular components get out of phase by the time it travels to 
position $z$, leading to the rotation of the polarization through the angle $\varphi$ 
given in expression (\ref{fr}).  One might also mention, that in general, the 
dielectric tensor elements are complex, then there is also a change in ellipticity 
${\cal X}$ of the polarization, given from the imaginary part,
\be
{\cal X} = \tfrac{1}{2} {\rm Im} \left( k_R-k_L \right) z \ .
\ee
The two effects of Faraday rotation and change in ellipticity ($\tan{\cal X}=$ ratio 
of minor to major axis of the ellipse swept out by the electric vector) can be 
combined into one complex parameter,\cite{sign_convention}
\be
\label{psi}
\psi = \varphi+i{\cal X} = \tfrac{1}{2} \left( k_R-k_L \right) z \ .
\ee
Usually these effects are extremely small and close to linear in ${\bf B}$.
Then there is only a tiny difference in $k_R$ and $k_L$, which gives to a very
good approximation, the complex relation,
\be
\label{psi1}
\psi = \varphi+i{\cal X} \approx \frac{\omega}{2c}\sqrt{\frac{\mu}{\epsilon_{xx}}}\, {\cal E}_{xy}\, z \ .
\ee
This emphasizes how the components of $\tilde\epsilon$ are needed to describe the
changes in the optical polarization.

From the experimental perspective, the measurement of the absorption (or, attenuation)
coefficient $\alpha$ is at least one technique that sets a relative scale for the FR.  
It is given from
\be
\label{alpha}
\alpha = 2\, {\rm Im} \left\{ k_{\rm eff} \right\} 
= 2\frac{\omega}{c} {\rm Im}\left\{ \sqrt{\mu\epsilon_{\rm eff}} \right\} \ .
\ee
This could use either $\epsilon_R$ or $\epsilon_L$ or their average for the effective 
dielectric function $\epsilon_{\rm eff}$ of the medium, as this expression does not 
involve their difference, which is extremely small.  Thus, measurements of $\alpha$ 
serve to set some unknown fitting parameters, when needed.

\section{Classical phenomenological model for \ew (Drude model)}
In this section the electron motion is assumed to be classical. An electron
of bare mass $m_o$ and charge $e=-1.602 \times 10^{-19}$ C has some 
trajectory ${\bf r}(t)=(x(t),y(t))$ in response to all forces acting on it, and
the averaging of its induced electric dipole moment ${\bf d}=e{\bf r}$
lead to the dielectric function.  

To include the effect of the constant ${\bf B}$ on $\tilde\epsilon$ it is assumed that there 
are two primary contributions to the dielectric response.  The first is the contribution 
of free electrons with number density $n$, and some damping parameter $\gamma_p$, 
that leads to the usual plasmon response with a plasma frequency 
$\omega_p^2=ne^2/m\epsilon_0$.  The second is a contribution due to bound electrons, 
with some binding frequency $\omega_0$ and another damping parameter $\gamma_0$.  
The contribution of bound electrons is essential to describe \ew correctly\cite{Scaffardi06} 
in NPs.

Any electron, whether free or bound, is acted on as well by the electric force 
from the optical field, and the Lorentz force from the DC magnetic field.  The 
force due to the optical magnetic field can be ignored in lowest order.  In this 
Drude approximation the equation of motion of a bound electron is\cite{Jackson}
\be
m_o\ddot{\bf r}=e{\bf E}+e\dot{\bf r}\times {\bf B}
-m_o\omega_0^2{\bf r}-m_o\gamma_0\dot{\bf r} \ .
\ee
Under the assumption of $e^{-i\omega t}$ time dependence of the optical field ${\rm E}$,
which is the source field, this is
\be
\left[m_o(\omega_0^2-\omega^2-i\omega\gamma_0)
-{i\omega e}{\bf B}\times \right] {\bf r} = e{\bf E} \ .
\ee
In terms of the components this is a matrix relation,
\be
\label{M}
\left[ \begin{array}{cc}
\omega_0^2-\omega^2-i\omega\gamma_0  & +i\omega \omega_{B} \\
-i\omega\omega_{B}  &  \omega_0^2-\omega^2-i\omega\gamma_0 \end{array} \right]
\left[ \begin{array}{c}  x  \\ y  \end{array} \right]
=\frac{e}{m_o} \left[ \begin{array}{c}  E_x  \\ E_y  \end{array} \right] \ ,
\ee
where the cyclotron frequency with ${\bf B}$ along $\hat{z}$ is
\be
\omega_B=\frac{eB}{m_o} \ .
\ee
The matrix $\tilde{\Omega}^2$ on the LHS of (\ref{M}) has the same kind of symmetry as 
that of $\tilde\epsilon$ in (\ref{eps-sym}), because the diagonal elements are equal 
and the imaginary off-diagonal elements differ only in sign.  This means $\tilde{\Omega}^2$
has the same eigenvectors, which are the right and left circular polarization states.
Based on its structure, the eigenvalues $\Omega^2_R$ and $\Omega^2_L$ of $\tilde{\Omega}^2$ 
are easy to read out. For right circular polarization,  
\be
\Omega^2_R = \omega_0^2-\omega^2-i\omega\gamma_0 +\omega \omega_{B} \ .
\ee
For left circular polarization, the last term (from the off-diagonal element) has the 
opposite sign,
\be
\Omega^2_L = \omega_0^2-\omega^2-i\omega\gamma_0 - \omega \omega_{B} \ .
\ee
The effect of the DC magnetic field appears only in the last factor.  These
two eigenvalues can be combined into a single convenient expression in terms of
the helicity $\nu=-1$ for right circular polarization and $\nu=+1$ for
left circular polarization:
\be
\Omega^2_{\nu} = \omega_0^2-\omega^2-i\omega\gamma_0 - \nu \omega \omega_{B} \ .
\ee
The helicity is the projection of the photon intrinsic angular momentum $\vec{L}$ on the 
direction of propagation (${\bf k}$ or $\hat{z}$).  In this expression it multiplies 
the magnetic field component along the same axis.  Any physical differences
for right and left circular polarizations will become interchanged if the 
direction of the magnetic field is reversed.  In terms of a
vector  $\vec\nu=\vec{L}/\hbar$, the last factor in these eigenvalues
could be written most generally as $\omega \vec{\nu}\cdot \vec{\omega}_B$.

If the electric field contains only one of the circular polarizations, i.e., 
${\bf E}=E_{\nu} \hat{\bf u}_{\nu}$, the response ${\bf r}$ also will be proportional
to the same eigenvector.  Then the solution for the electron position is very simple,
\be
{\bf r} = \frac{eE_{\nu}}{m_o\Omega^2_{\nu}}\hat{\bf u}_{\nu} \ .
\ee
The factor $e/m\Omega^2_{\nu}$ gives the size of the response for this polarization. 
Here we see the fundamental physical difference between the polarizations.  One of 
the polarizations causes a larger circular motion of the electrons than the other 
polarization.  Which one actually is larger depends on the relation between $\omega_0$ 
and $\omega$. This difference leads to a corresponding difference in the dielectric effects.

Based on this position response, it is then easy to find the effective
dielectric functions for the two circular polarizations, using ${\rm d}=e{\rm r}$.
The susceptibility due to these electrons is $\chi=n\langle {\rm d}\rangle / \epsilon_0 E$.
The result can be summarized in a single formula,
\be
\label{chinu}
\chi_{\nu}(\omega) = \frac{ne^2}{m_o\epsilon_0\, \Omega^2_{\nu}}\ ,
\ee
where $\nu=-1/+1$ refers to $R/L$ polarizations, respectively. This applies to
separately, the contribution from the bound electrons, or, the contribution of
the free electrons, using appropriate parameters in each case.

Look at this another way.
An arbitrary electric field can be expressed either as
${\bf E}=E_x \hat{x}+E_y \hat{y}$ or as  
${\bf E}=E_R \hat{\bf u}_R +E_L \hat{\bf u}_L$, where
\bn
E_R &=& \tfrac{1}{\sqrt{2}}\left(E_x+iE_y\right), \quad
E_x = \tfrac{1}{\sqrt{2}}\left(E_R+E_L\right), \\
E_L &=& \tfrac{1}{\sqrt{2}}\left(E_x-iE_y\right), \quad
E_y = \tfrac{-i}{\sqrt{2}}\left(E_R-E_L\right).
\en
One can combine the right and left solutions and get the general solution
for any electric field, in diagonal form:
\be
{\bf r}=\frac{eE_R}{m_o\Omega^2_R}\hat{\bf u}_R +\frac{eE_L}{m_o\Omega^2_L}\hat{\bf u}_L \ .
\ee
Alternatively, this can be written in Cartesian components,
\bn
x &=& \frac{1}{\sqrt{2}}\frac{e}{m_o}\left[ \frac{E_R}{\Omega^2_R}+\frac{E_L}{\Omega^2_L} \right]\ , \\
y &=& \frac{1}{\sqrt{2}}\frac{ie}{m_o}\left[ \frac{E_R}{\Omega^2_R}-\frac{E_L}{\Omega^2_L}\right]\ .
\en
Simplifying, or inverting the matrix equation (\ref{M}), leads to the
general electron motion,
\be
\left[ \begin{array}{c}  x  \\ y  \end{array} \right] =
\frac{{e}/{m_o}}{\Omega^2_R\Omega^2_L}
\left[ \begin{array}{cc} \omega_0^2-\omega^2-i\omega \gamma_0 & -i\omega\omega_B \\
+i\omega\omega_B  & \omega_0^2-\omega^2-i\omega\gamma_0 \end{array} \right] 
\left[ \begin{array}{c}  E_x  \\ E_y  \end{array} \right] \ .
\ee
Multiplied by $e$, the relation shows the polarizability matrix of the electron.  
This expression leads to the susceptibility tensor,
\be
\tilde\chi = \frac{ne^2/m_o\epsilon_0}{\Omega^2_L\Omega^2_R}
\left[ \begin{array}{cc} \omega_0^2-\omega^2-i\omega \gamma_0 & -i\omega\omega_B \\
+i\omega\omega_B  & \omega_0^2-\omega^2+i\omega\gamma_0 \end{array} \right] \ .
\ee
One can see this is consistent with (\ref{chinu}), because its eigenvalues are
$\chi_R=\chi_{xx}-i\chi_{xy}$ and $\chi_L=\chi_{xx}+i\chi_{xy}$, which agrees
exactly with (\ref{chinu}).

\subsection{Combination of free and bound electron responses}
Now to use this to describe a metal such as gold, we assume first there is some density of
free electrons $n$, with a bulk plasma frequency $\omega_p^2=ne^2/m_o\epsilon_0$, a damping
$\gamma_p$ and a zero binding frequency.  In addition, there is some other density $n_0$ of 
bound electrons, leading to an effective weight $g_0^2=n_0 e^2/m_o\epsilon_0$, with an
associated damping $\gamma_0$ and binding frequency $\omega_0$.  The net dielectric
function is the sum of the two contributions to $\tilde\chi$.  In terms of
the polarization states $\nu = \pm 1$, the dielectric function is taken as
\be
\label{epsCM}
\epsilon_{\nu}(\omega) = 1-\frac{\omega_p^2}{\omega^2+i\omega\gamma_p+\nu\omega\omega_B}
-\frac{g_0^2}{\omega^2-\omega_0^2+i\omega\gamma_0+\nu\omega\omega_B} \ .
\ee
The first two terms are the usual ones for describing a free electron gas. The
last term uses the single resonance to approximate the effects of bound electrons.
Both include the DC magnetic field implicitly in the cyclotron frequency, $\omega_B$.
The ease with which the magnetic field is included in the bound electron response is
the main advantage of this model.

One can produce the Cartesian elements of $\tilde\epsilon$, for instance,
using (\ref{epsR}) and (\ref{epsL}), by the combinations of these eigenvalues:
\bn
\epsilon_{xx} &=& \epsilon_{yy}=\tfrac{1}{2}\left( \epsilon_R+\epsilon_L \right) \ , \\ 
\epsilon_{xy} &=& -\epsilon_{yx}=i {\cal E}_{xy} = \tfrac{i}{2}\left( \epsilon_R-\epsilon_L \right) \ . 
\en

\subsection{Maxwell-Garnett averaging for dilute solutions}
The medium of interest here is actually a dilute solution of NPs at a volume
fraction $f_s \ll 1$ in a host liquid, which we take as water, with its host 
dielectric constant $\epsilon_h=1.777$.  The NPs are considered the scatterers
with dielectric function $\epsilon_s$ 
For comparison with experiment, the 
effective dielectric function $\epsilon_{\rm eff}$ of the solution is required. 
The theory for calculating the effective dielectric function depends somewhat on
the assumption of how the particles are dispersed in the liquid.  In the simplest 
approximation, they are assumed to be randomly dispersed and not sticking to each 
other.  In the Maxwell-Garnett (MG) theory,\cite{MG04,BW65} one finds the volume-averaged 
electric field and the volume-averaged polarization response to that field, from which 
$\epsilon_{\rm eff}$ is determined.  The MG theory is known to apply well even in the
presence of multiple-scattering.\cite{Mallet+05} These volume averages are
\bn 
{\bf E}_{\rm av} &=& f_s {\bf E}_s +(1-f_s) {\bf E}_h \ , \\
{\bf P}_{\rm av} &=& f_s {\bf P}_s +(1-f_s) {\rm P}_h \ ,
\en
where $s$ and $h$ refer to the values in the scatterer and the host, respectively.
For spherical scatterers exposed to asymptotic field ${\bf E}_h$ in the host, the 
Clausius-Mosotti equation gives the internal fields,
\be
{\bf E}_s = \frac{3\epsilon_h}{\epsilon_s+2\epsilon_h} {\bf E}_h \ ,
\quad
{\bf P}_s = \left( \epsilon_s-1 \right) \epsilon_0 {\bf E}_s \ .
\ee
Then with polarization ${\bf P}_h = (\epsilon_h-1) \epsilon_0 {\bf E}_h$ in the host,
one finds the average
\be
\label{epseff}
\epsilon_{\rm eff} = 1+\frac{{\bf P}_{\rm av}}{\epsilon_0{\bf E}_{\rm av}} 
= \epsilon_h \frac{1+2\beta_f}{1-\beta_f}\ ,
\ee
which involves the scaled volume fraction ($f_s$ is the fraction of volume occupied
by NPs in the solution),
\be
\beta_f = f_s  \frac{\epsilon_s-\epsilon_h}{\epsilon_s+2\epsilon_h} \ .
\ee
This MG averaging procedure for composite systems is usually summarized by the equivalent 
relation,
\be
\frac{\epsilon_{\rm eff}-\epsilon_h}{\epsilon_{\rm eff}+2\epsilon_h} =
f_s \frac{\epsilon_s-\epsilon_h}{\epsilon_s+2\epsilon_h} \ .
\ee
Expression (\ref{epseff}) can be applied separately to the left and right circular
polarization states, then leading to an effective dielectric function for each,
that will then give the Faraday rotation (\ref{fr}) for a dilute solution.

\subsection{Classical model parameters for gold nanoparticles}
\begin{figure}
\begin{center}
\includegraphics[width=\smallfigwidth,angle=-90]{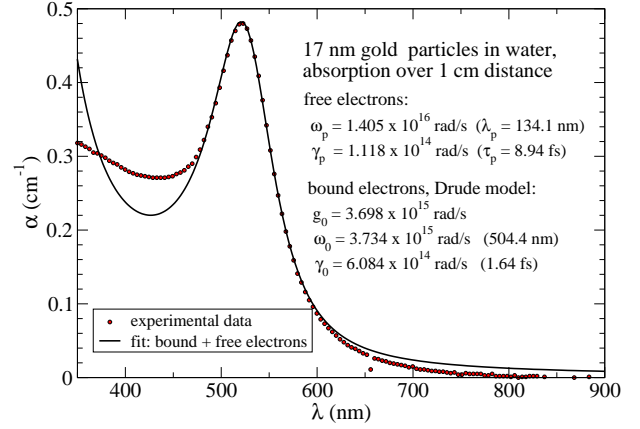}
\end{center}
\caption{\label{gold-abs-drude} (Color online) Fitting of the absorption of 17 nm diameter gold particles 
in water solution, according to the Drude model for the bound electrons.  Parameters indicated are 
used to get a good fit to the absorption peak near 522 nm. The fitted volume fraction of gold 
is $f_s=3.36 \times 10^{-6}$.} 
\end{figure}

\begin{figure}
\begin{center}
\includegraphics[width=\smallfigwidth,angle=-90]{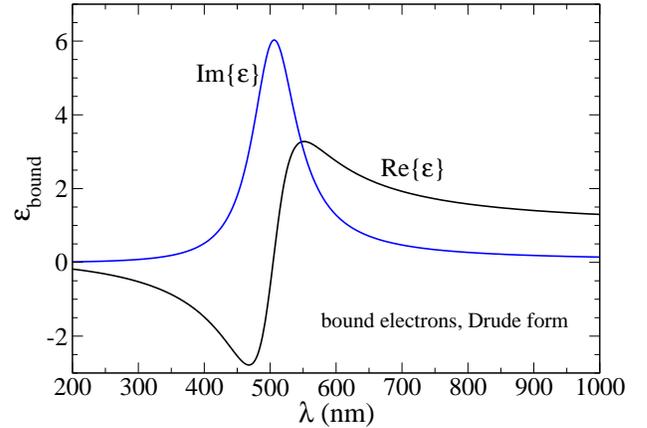}
\end{center}
\caption{\label{eps-bnd-drude} (Color online) The bound electron contribution to the permittivity, from the
second term of Eq.\ (\ref{epsCM}), according to the Drude model for the bound electrons, using the 
parameters of Figure \ref{gold-abs-drude}.  The real part of \ew becomes negative for frequencies 
above $\omega_0$ (wavelength 504 nm), which is a defect of this model.}
\end{figure}

Based on the work in Ref. \onlinecite{Dani11}, the parameters needed for this classical
model were found by fitting it to the absorption measured experimentally with $B=0$, 
for a dilute solution of 17 nm diameter gold NPs in water. 
That fitting is based on using the effective dielectric function $\epsilon_{\rm eff}$ 
from the MG theory, to give the absorption in the solution, according to  expression
(\ref{alpha}).

For this classical Drude model, based on the electron number density, and using
effective mass equal to the bare electron mass, the bulk plasma frequency is 
$\omega_p = 1.37 \times 10^{16}$ rad/s, which corresponds to $\lambda_p=2\pi c/\omega_p=138.5$ nm.  
The damping of the free electrons in NPs can have an intrinsic term and a surface scattering term.  
Thus a size-dependent damping factor is included, according to the combination of these processes,
\cite{Antoine++97}
\be
\label{gp}
\gamma_p = \frac{1}{\tau}+\frac{v_F}{d} \ ,
\ee
where $\tau\approx 9.1$ fs is the intrinsic scattering time, $v_F = 1.40\times 10^6$ m/s 
is the Fermi velocity, and $d$ is the thickness of the gold. This thickness could be the 
diameter for solid spherical particles, or, the thickness of a shell for core/shell particles. 
We discuss data for gold particles of average diameter 17 nm; the prediction for their effective
damping is then $\gamma_p= 1.92 \times 10^{14}$ rad/s, or, a time scale $\tau_p=\gamma_p^{-1}=5.20$ fs.

The Drude theory was fitted to experimental data for absorption through a 1 cm path of water solution
of gold particles with average diameter of 17 nm.  The fitting parameters were chosen
to get a good description of the absorption peak present near 522 nm, attributed to 
surface plasmon response.  A good description can be obtained while also allowing the volume 
fraction and free electron parameters $\omega_p$ and $\gamma_p$ to vary,
see Figure \ref{gold-abs-drude}.  The contribution from the bound electrons can 
be represented approximately using the amplitude parameter $g_0=3.70 \times 10^{15}$ rad/s, 
the binding frequency $\omega_0 = 3.73 \times 10^{15}$ rad/s (wavelength 504 nm), and damping 
frequency $\gamma_0 = 6.08 \times 10^{15}$ rad/s, which corresponds to a damping time of 
$\tau_0 =1/\gamma_0 \approx 1.64$ fs.  To get this good fit to the peak, the free electrons
are at the same time represented using plasma frequency $\omega_p=1.40 \times 10^{16}$ rad/s, 
equivalent to $\lambda_p=134.1$ nm, and a damping $\gamma_p=1.118 \times 10^{14}$ rad/s, corresponding 
to the damping time $\tau_p=8.94$ fs.  These are slightly different than the accepted bulk values, 
however, we consider them here only as a model that fits accurately the absorption peak.  

\begin{figure}
\begin{center}
\includegraphics[width=\smallfigwidth,angle=-90]{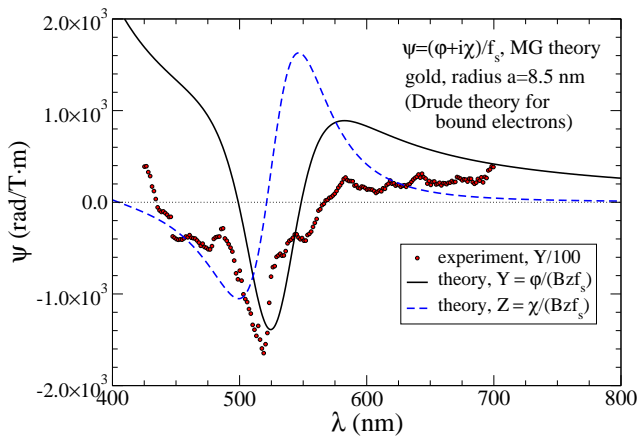}
\end{center}
\caption{\label{psi-gold17-drude} (Color online) Faraday rotation for 17 nm diameter gold
NPs, from experiment, and according to the Drude model for the bound electrons, using the 
Drude fitting parameters of Figure \ref{gold-abs-drude}.  The experimental results have been
scaled by $1/100$ to allow them to be plotted together with the Drude theory. The theory result 
was obtained with the MG effective medium approach.  Faraday rotation angle $\varphi$ and 
ellipticity angle ${\cal X}$ have been scaled by $B, z, f_s$, to give $\Psi = \Upsilon + i Z$.}
\end{figure}

From these fitted dielectric parameters, the theoretical Faraday rotation response can be obtained.  
Results for the Faraday rotation and ellipticity for 17 nm gold NPs in solution are shown in 
Fig.\ \ref{psi-gold17-drude}.  The complex rotation angle $\psi$ is found from Eq.\ \ref{psi},
together with applying the Maxwell-Garnett procedure for the composite medium, Eq.\ \ref{epseff}, 
for the effective dielectric function of the composite solution. We have
scaled the rotation angle $\varphi$ and ellipticity ${\cal X}$ by the product of path length $z$,
magnetic field $B$ and gold volume fraction $f_s$, to remove the linear dependence on these quantities.
Thus we define the complex rotation angle scaled by volume fraction,
\be
\Psi \equiv \Upsilon + i Z = \psi/f_s \ .
\ee
Then, $\Upsilon \equiv \varphi/(Bzf_s)$  is the Verdet factor per unit volume fraction, and 
$Z\equiv {\cal X}/(Bzf_s)$ is a corresponding ellipticity factor per unit volume fraction.  
Then the results for $\Upsilon$ and $Z$ do not depend on $B$, $z$, or $f_s$ in the linear regime.  
The experimental data for $\Upsilon$ are also displayed in Fig.\ \ref{psi-gold17-drude},
scaled down by a factor of $1/100$ in order to be shown together with the theory.

One sees that the model predicts a negative peak in the Faraday rotation near 525 nm, 
apparently associated with the plasmon resonance
(see \onlinecite{sign_convention} for the distinction between positive and negative rotation
angles). The experimental data have a similar negative peak in the same region, although its
magnitude is significantly larger than this theory predicts.   The theory has a wider positive 
peak around 580 nm and a long tail at longer wavelengths, but this positive peak is rather
weak in the experimental data.  For the ellipticity, the main feature predicted is a positive peak 
around 540 nm, slightly
above the plasmon wavelength, together with its associated negative peak and long tail at shorter
wavelengths.  Unfortunately, the model exhibits an artifact at shorter wavelengths: both $\varphi$
and ${\cal X}$ tend to increase greatly at short wavelengths in an unphysical behavior. This is 
due to the fact that classical Drude model cannot correctly describe the bound electron response at 
higher frequencies.

This model is an approximate way to include the effect of $B$ on classical bound electrons, 
however, it should be replaced by the more complete calculation using the quantum interband
transitions presented later.  It gives a reasonable fit to the absorption curve from 900 nm down 
to 400 nm, however, below that wavelength it predicts much more absorption than actually takes place.
Also, this Drude description of the bound electrons cannot accurately describe the response 
in the wavelengths 350 -- 500 nm.  This model does not require any background (i.e., high-frequency) 
dielectric function $\epsilon_{\infty}\sim 10$, as has been applied in other studies to mimic the 
effect of bound electrons.\cite{Antoine++97}  Even so, the fit to the absorption peak due to the 
SP mode is very good, while the corresponding negative FR peak due to the SP mode of 17 nm gold 
particles is about 100 times stronger than the theory predicts.

\section{Quantum description of \ew via perturbation of the density matrix}
In this Section we consider the quantum calculation of the effects due to bound
electrons, which is taken into account by finding contributions to \ew due to interband
transitions, in the presence of the DC magnetic field.  The electrons are considered
non-interacting.

The single-electron Hamiltonian is taken as 
\be
\label{Helectron}
\hat{H}=\frac{1}{2m_o}\left[\hat{\bf p}-{e}\hat{\bf A}(\hat{\bf r},t)\right]^2
+e\hat\phi(\hat{\bf r},t) +\hat{U}(\hat{\bf r}) \ ,
\ee
where the charge is $e$, $\hat\phi$ and $\hat{\bf A}$ are the scalar and vector potentials 
of the EM fields, and $\hat{U}$ is the periodic potential of the lattice. 
The canonical momentum operator for the electron is $\hat{\bf p}=-i\hbar {\vec\nabla}$. 
The more physical momentum is the kinetic momentum operator,
\be
\vec\pi = \hat{\bf p}-{e} \hat{\bf A} \ ,
\ee
because it is the square of this operator that determines the energy. We take the scalar 
potential as zero (Coulomb gauge).   The vector potential $\hat{\bf A}$ includes 
a term for the DC magnetic field, $\hat{\bf A}_0$ and a term for the AC optical field $\hat{\bf A}_1$.
The optical field is treated as a classical non-quantized field, that oscillates as $e^{-i\omega t}$.

The electron bands, unperturbed by optical fields, come from the solution of a Hamiltonian
with the kinetic energy,  the lattice periodic potential, and the DC magnetic field,
\be
\hat{H}_0 = \frac{1}{2m_o}\left(\hat{\bf p}-{e}\hat{\bf A}_0\right)^2+U({\bf r}) \ .
\ee
At weak enough DC magnetic field, the quadratic term in $\hat{\bf A}_0$ can be dropped, and
the effect of the cross term with $\hat{\bf p}$ is the orbital Zeeman splitting,
\be
\hat{H}_0 = \frac{\hat{\bf p}^2}{2m_o}+U({\bf r})-\vec{\mu}\cdot {\bf B} \ .
\ee
The magnetic dipole moment due to the orbital angular momentum is
\be
\vec{\mu} = \frac{e}{2m_o} \vec{L} \ .
\ee
As the electron charge is negative, $\vec\mu$ points opposite to $\vec{L}$.  The component 
of $\vec{\mu}$ along the magnetic field is $m \mu_B$, where $\mu_B=e\hbar/2m_o$ is the 
(negative) Bohr magneton, and $m=m_l$ is the magnetic quantum number. The Zeeman splitting 
is an energy shift $\Delta E = - m \mu_B B = - \tfrac{1}{2}m \hbar\omega_B$,  
where both $\mu_B$ and $\omega_B$ can be negative, due to the negative electron charge.  
The states of this Hamiltonian are some electron band states, including any Zeeman shifts,
\be
\hat{H}_0 |{\bf k}lm\rangle = E_{{\bf k}lm} |{\bf k}lm\rangle \ .
\ee
The band states, labeled by wave vector ${\bf k}$ and angular indexes $l,m$, have
wave functions
\be
\psi_{{\bf k}lm}({\bf r})=
\langle {\bf r} |{\bf k}lm\rangle = \frac{1}{\sqrt{V}} e^{i \bf k\cdot r} u_{{\bf k}lm}({\bf r}) \ .
\ee
These can be considered the original states of an unperturbed problem.  The optical field is 
the perturbation on these states, whose effect is studied using the density matrix approach.  

Because we consider states in NPs, the DC magnetic field only produces Zeeman shifts, rather than 
Landau levels.  Due to the geometrical confinement, there is no sense to Landau levels that would 
have extended wave functions much larger than the size of the particles.  For instance, at a small 
applied field strength $B=0.1$ tesla,  the length scale of the Landau levels is the 
Landau radius, $r_0=\sqrt{\tfrac{2\hbar}{eB}} = 115$ nm.  This is much larger than the
radius of the nanoparticles under consideration, typically from 5 -- 10 nm.  The Landau wave functions
do not fit into the NPs at this field strength, giving a non-bulk situation.  The degeneracy 
of Landau levels is on the order of $(R/r_0)^2$, where $R$ is the system radius.  At $B=0.1$ tesla,
the degeneracy is about $(8.5/115)^2\approx 0.0055$, however, this fractional value is not
meaningful.  For the larger magnetic field $B=4.2$ T, the Landau radius is reduced to $r_0=17.7$ 
nm.  This is still somewhat larger that the NP radius of 8.5 nm, and the degeneracy is about
$(8.5/17.7)^2 \approx 0.23$, still significantly less than 1, so the theory should be applicable.

These considerations show that the Landau levels are the incorrect solutions in a confined geometry.  
When one looks more carefully at how to arrive at the quantum solution, the radial wave functions 
should go to zero at the boundary of the NP (for bound electrons).  For spherical particles, that
radial dependence would be described by spherical Bessel functions, $j_l(kr)$, with discrete
allowed $k$, and angular dependence described by spherical harmonics for a chosen angular momentum, 
$l,m$.  We consider a quasi-bulk approximation, where the discrete $k$ are assumed to be
close enough together to be reasonably described by electron bands.

\subsection{The density operator $\hat{\rho}$}
Statistically, the band states are populated according to a Fermi-Dirac distribution for the
given temperature, when the system is in equilibrium.  The density operator $\hat\rho$ is a
way to introduce this population into the QM problem and provide for mixed states.
Of course, once the optical field is turned on, a new equilibrium can be established 
and the density operator can change.  Its basic definition for an equilibrium situation, in 
terms of the state probabilities $w_i$ is
\be
\hat\rho_0 = \sum_i w_i |\psi_i\rangle \langle \psi_i | \ .
\ee
For the equilibrium distribution, the weights are taken as proportional to the Fermi-Dirac
occupation numbers,
\be
w_i = \frac{1}{N}f_0(E_i), \quad f_0(E_i) = \frac{1}{e^{\beta(E_i-E_F)}+1} \ ,
\ee
where $\beta$ is the inverse temperature and $E_F$ is the Fermi level.
This is a density matrix normalized to one for $N$ electrons in the system.
The time derivative of a general $\hat{\rho}$ follows from the quantum Liouville equation:
\be
\frac{\partial \hat\rho}{\partial t} =\frac{1}{i\hbar} [\hat{H},\hat\rho] \ .
\ee

To apply this, we consider the leading perturbation term in the Hamiltonian,
which is the electric force from the optical field ($\hat{\bf A}_1$). The optical 
magnetic force is ignored.  Then the perturbation is described by the Hamiltonian
\be
\hat{H}_1 =  -\frac{e}{m_o}\hat{\bf A}_1 \cdot \left(\hat{\bf p}-{e}\hat{\bf A}_0\right) \ .
\ee
Now the total density operator is assumed to be a sum of the equilibrium operator
plus some change caused by the perturbation:
\be
\hat{\rho} = \hat{\rho}_0 + \hat{\rho}_1 \ .
\ee 
As the total Hamiltonian also is a sum of unperturbed and perturbation parts, we can use the 
fact that $[\hat{H}_0,\hat{\rho}_0]=0$, and ignore the small nonlinear term 
$[\hat{H}_1,\hat{\rho}_1] \approx 0$, then the equation of motion for the perturbation is
\be
i\hbar  \frac{\partial \hat\rho_1}{\partial t} \approx  
[\hat{H}_0,\hat\rho_1]+[\hat{H}_1,\hat\rho_0] \ .
\ee
Now assume expansions of $\hat{H}_1$ and of $\hat\rho_1$ in the unperturbed basis states, 
\be
\hat{H}_1 = \sum_{if} |f\rangle \langle f| \hat{H}_1 |i\rangle \langle i|, \hskip 0.2in
\hat\rho_1 = \sum_{if} c_{f i} |f\rangle  \langle i|.
\ee
The constants $c_{fi}=\langle f | \hat{\rho}_1 | i \rangle$ are just
the matrix elements of $\hat\rho_1$, in the $H_0$ basis states.  After evaluation of the
commutators, and assuming $e^{-i\omega t}$ time dependence for $\hat{\rho}_1$, the constants
$c_{fi}$ are found, and the change in the density operator is found to be
\be
\label{cfi}
\hat\rho_1 = \sum_{if} \frac{(w_i-w_{f}) | f\rangle \langle f| \hat{H}_1 | i \rangle \langle i| }
{\hbar(\omega+i\gamma) +(E_i-E_{f}) }\ .
\ee
A small imaginary part $\gamma$ has been added to the frequency to effect the turning on of the 
perturbation.  This constant can be considered a phenomenological damping constant, or, it can 
be let to go to zero if the results without damping are of interest.  This expression has been used
in various problems by Adler\cite{Adler62} and in the thesis of M.\ Prange.\cite{Prange09}  
$E_i$ and $E_{f}$ are energies of two states of the unperturbed  Hamiltonian.  One can think that 
the expression involves transitions between pairs of states. Obviously the oscillatory time behavior 
of the perturbation Hamiltonian must be reflected in a similar behavior in this part of the density matrix.  
Thus, we are interested only in the response in the density matrix at the same frequency as the 
perturbation.

\subsection{Thermal and volume averages}
To find the dielectric function, statistical averages of the polarization or the current 
density are necessary.  This can be done by first defining a local quantum operator, 
$\hat{\rho}_e({\rm r})$ for the one-electron charge density, 
\be
\hat{\rho}_e({\rm r}) = e\vert {\bf r} \rangle \langle  {\bf r} \vert \ ,
\ee
and another, $\hat{j}({\bf r})$, for the one-electron current density,
\be
\hat{j}({\bf r}) = \frac{e}{2} \left\{ \vert {\bf r} \rangle \langle  {\bf r} \vert \hat{\bf v}
+ \hat{\bf v} \vert {\bf r} \rangle \langle {\bf r} \vert \right\} \ .
\ee
The current density operator is defined in terms of the electron velocity,
\be
\hat{\bf v}=\frac{\vec{\pi}}{m_o}=\frac{1}{m_o}\left( \hat{\bf p} -{e} \hat{\bf A}\right) \ .
\ee
The statistically averaged values of these operators are found from the trace with the density
operator,
\bn
\rho_e({\rm r}) &=& \langle \hat{\rho}_e({\bf r}) \rangle 
= {\rm Tr}\left\{ \hat{\rho} \hat{\rho}_e({\bf r}) \right\} \ , \\
{\bf j}({\rm r}) &=&  \langle \hat{j}({\bf r}) \rangle 
= {\rm Tr}\left\{ \hat{\rho} \hat{j}({\bf r}) \right\} \ . 
\en
In a pure state $\vert \psi \rangle$, with density operator 
$\hat\rho=\vert \psi \rangle \langle \psi \vert$, these produce the usual expressions for the 
quantum charge and current densities at point ${\bf r}$,
\bn
\rho_e({\rm r}) &=& e\vert \psi({\bf r}) \vert^2 \ , \\
{\bf j}({\bf r}) &=& {\rm Re}\left\{ \psi^*({\bf r}) e\hat{\bf v} \psi({\bf r}) \right\} \ .
\en
We can also define the local polarization operator using the electron position,
\be
\hat{\bf d}({\bf r}) = e \hat{\bf r} \vert {\bf r} \rangle \langle {\bf r} \vert \ ,
\ee
which is statistically averaged by the same procedure,
\be
{\bf d}({\bf r}) = {\rm Tr}\left\{ \hat{\rho} \hat{\bf d}({\bf r}) \right\} \ .
\ee

At some point in the calculation the volume averages are desired, to describe \ew
for the whole sample.  These are obtained from the usual definition, say, for the
charge density (the overbar indicates volume average) due to $N$ electrons
in a volume $V$,
\be
\overline{\rho}_e = \frac{N}{V}\int d^3r\, \rho_e({\rm r})  
=\frac{Ne}{V} \sum_i w_i \int d^3r\, \vert\psi_i({\bf r})\vert^2 \ .
\ee
The individual electron states $\psi_i$ are unit normalized and the probabilities
sum to one.  This recovers an obvious result, 
\be
\rho = N \overline{\rho}_e = \frac{N}{V} {\rm Tr}\left\{ \hat{\rho}e \right\} = ne \ .
\ee
For the current density and the electric polarization, there are similar expressions,
\bn
{\bf J} = N \overline{\bf j} &=& \frac{N }{V}\int d^3r\, {\bf j}({\bf r})
= {\rm Tr}\left\{ \hat{\rho} ne\hat{\bf v} \right\} \ , \\
{\bf P} = N \overline{\bf d} &=& \frac{N}{V}\int d^3r\, {\bf d}({\bf r})
=  {\rm Tr}\left\{ \hat{\rho}  ne\hat{\bf r} \right\} \ .
\en
For the most part, we use the averaging of the operator,
\be
 \hat{\cal P}= ne\hat{\bf r} \ ,
\ee
to determine the volume-averaged polarization response.

\subsection{Averaged electric polarization response}
The perturbation oscillates at frequency $\omega$ and therefore we get the 
response in the electric polarization ${\bf P}$ that oscillates at the same 
frequency, using only the change $\hat{\rho}_1$ in the density matrix,
\be
{\bf P} = {\rm Tr}\left\{ \hat{\rho}_1 \hat{\cal P} \right\} 
= ne \sum_{if} \langle f \vert \hat{\rho}_1 \vert i \rangle 
\langle i \vert \hat{\bf r} \vert  f \rangle \ .
\ee
This requires matrix elements of the position operator in the unperturbed basis states.
Those can be obtained from the equation of motion in the unperturbed system,
\be
i\hbar \dot{ {\bf \hat{r}} }= i \hbar \hat{\bf v} = [\hat{\bf r}, \hat{H}_0 ]\ .
\ee
Then the needed matrix elements can be expressed using the velocity,
\be
\label{rmes}
\langle i \vert \hat{\bf r} \vert f \rangle 
= \frac{i\hbar}{(E_{f}-E_{i})} \langle i \vert \hat{\bf v} \vert f \rangle \ .
\ee
The optical electric field is ${\bf E} = -(\partial \hat{\bf A}_1/\partial t )= 
i(\omega+i\gamma)\hat{\bf A}_1$, including the turning on of the perturbation.  
The perturbation can be expressed now as
\be 
\hat{H}_1 =  -e\hat{\bf A}_1 \cdot \hat{\bf v} 
= \frac{-e}{i(\omega+i\gamma)} {\bf E}\cdot\hat{\bf v} \ ,
\ee
The operator $\hat{\bf v}$ need include only the DC vector potential, $\hat{\bf A}_0$. 
Then the matrix elements of both $\hat{\rho}$ and $\hat{H}_1$ come from the velocity. 
The result for the averaged electric polarization is expressed as
\be
\label{P1}
{\bf P}=\frac{ne^2\hbar}{(\omega+i\gamma)} \sum_{if}
\frac{(w_i-w_{f}) \langle i\vert \hat{\bf v}\,\vert f\rangle 
                   \langle f\vert {\bf E}\cdot\hat{\bf v}\,\vert i\rangle }
{\left[\hbar(\omega+i\gamma)+E_i-E_{f}\right] \left(E_i-E_{f}\right)}  \ .
\ee
As ${\bf E}$ is assumed to oscillate at frequency $\omega$, this is indeed the
response oscillating at that same frequency. The damping $\gamma$ is necessary 
so that an appropriate limit  gives the classical damped responses found earlier.

With the transition frequencies given by
\be
\hbar \omega_{if}= E_i-E_{f} \ ,
\ee
the susceptibility components that result from (\ref{P1}) are
\be
\label{chiij}
\chi_{ab}=
\frac{ne^2}{\epsilon_0\hbar(\omega+i\gamma)} \sum_{if}
\frac{(w_i-w_{f}) \langle i\vert \hat{v}_{a}\vert f\rangle 
                   \langle f\vert \hat{v}_{b}\vert i\rangle }
{ \omega_{if} \left(\omega+i\gamma+\omega_{if}\right) } \ .
\ee
We apply this to find only the effects from interband transitions.  
The free electron response in (\ref{epsCM}) is still applied for the quantum model.

 \begin{widetext}
 The result can be symmetrized by labeling some states as occupied states
 (o) and the rest as unoccupied (u). All terms correspond to transitions
 from occupied to unoccupied states. In this way the expression becomes
 \be
 \chi_{ab}=
 \frac{ne^2}{\epsilon_0\hbar(\omega+i\gamma)} \sum_{i}^{o} \sum_{f}^{u}
 \frac{w_i-w_{f}}{\omega_{if}} \left\{
 \frac{ \langle i\vert \hat{v}_{a}\vert f\rangle \langle f\vert \hat{v}_{b}\vert i\rangle }
 { \omega+i\gamma+\omega_{if} }
 +\frac{ \langle i\vert \hat{v}_{b}\vert f\rangle \langle f\vert \hat{v}_{a}\vert i\rangle }
 { \omega+i\gamma-\omega_{if} } \right\}
 \ee
 \end{widetext}

\subsection{Application to band models}
To apply this result, we need to use the energy levels appropriate for the bands
under consideration.  The discussion is restricted to a parabolic two-band model, with the
bands separated by some gap energy $E_g$.  There are effective masses $m^*_h$ and $m^*_e$ for 
the occupied (lower) and unoccupied (higher) bands, respectively.  Each band is affected by the 
Zeeman shift in the same direction; there are not Landau level shifts. One can measure energies 
from the top of the lower band.  Then the energies $E_i$ for the occupied band ($E_h$, valence band) 
and the energies $E_f$ for the unoccupied band ($E_e$, conduction band) are assumed to be
\bn
E_i &=& E_h =  -\frac{\hbar^2 {\bf k}_i^2}{2m^*_h}-\tfrac{1}{2} m_i \hbar\omega_B \ , \\
E_f &=& E_e = E_g+\frac{\hbar^2 {\bf k}_f^2}{2m^*_e}-\tfrac{1}{2} m_f \hbar\omega_B \ , \\
\en
These Zeeman shifts apply to positive charges; they are reversed in sign for negative charges,
taking $\omega_B<0$.  The azimuthal quantum numbers are $m_i$ and $m_f$.  They are restricted 
by the orbital angular momentum numbers for each band, $l_i$ and $l_f$, respectively.  Assuming 
vertical transitions that conserve linear momentum $\hbar {\bf k}$ (negligible photon momentum), 
the transition energies are
\be
\hbar\omega_{if} = 
-  E_g - \frac{\hbar^2 {\bf k}^2}{2m^*} - \tfrac{1}{2} (m_i-m_f)\hbar\omega_B \ ,
\ee
where the reduced mass $m^*$ is defined by
\be
\frac{1}{m^*}=\frac{1}{m^*_e}+\frac{1}{m^*_h} \ .
\ee
We write the transition frequencies in the following manner:
\be
\omega_{if} = - \omega_g - s^2 + \tfrac{1}{2}\Delta m\, \omega_B \ , 
\ee
where the gap frequency $\omega_g$, scaled wave vector $s$, and  
 change in azimuthal quantum number $\Delta m$ are
\be
\omega_g \equiv \frac{E_g}{\hbar}, \quad 
s \equiv \sqrt{\frac{\hbar}{2m^*}}\, k, \quad 
\Delta m \equiv m_f-m_i.
\ee

Only momentum-conserving transitions between two selected bands at some wave vector ${\bf k}$ 
are considered. The matrix elements needed are approximated in a form 
\be
\langle {\bf k'}l'm' \vert \hat{v}_x \vert {\bf k}lm \rangle 
= \frac{\hbar k_x}{m_o} M({\bf k}) \delta_{\bf k', k}\delta_{l', l\pm 1}\delta_{m', m\pm 1} \ .
\ee
The last Kronecker deltas reflect the electric dipole selection rules, $\Delta l=\pm 1$,
$\Delta m =\pm 1$.  The dimensionless matrix element $M({\bf k})$ is assumed to
be some constant for the transitions of interest.

These velocity matrix elements are proportional to corresponding position matrix elements,
see (\ref{rmes}), or even the matrix elements of the $\vec{\pi}$ operator.  We only need the 
components of operators along $x$ and $y$.  But the angular part of these matrix elements is 
due to the electric dipole selection rules.  That angular part has the following symmetries,
from matrix elements between spherical harmonics,
\be
\langle l'm' \vert \hat{v}_y \vert l m \rangle 
= -i \Delta m\, \langle l'm' \vert \hat{v}_x \vert lm \rangle  ,
\quad
\Delta m = \pm 1 \ .
\ee
This directly affects the susceptibility for each circular polarization.  From (\ref{chiij})
we have the diagonal part as
\be 
\chi_{xx} \sim \sum_{fi} g_{fi} \left\vert \langle f \vert \hat{v}_x \vert i \rangle \right\vert^2 \ ,
\ee
but the off-diagonal part as
\be 
\chi_{xy} \sim \sum_{fi} \left(-i\Delta m\right) g_{fi} 
\left\vert \langle f \vert \hat{v}_x \vert i \rangle \right\vert^2 \ .
\ee
It is clear that $\tilde\chi$ and $\tilde\epsilon$ have the same symmetry.
Then the susceptibilities for the right and left circular polarizations
vary like
\bn
\chi_R &=&  \chi_{xx}-i\chi_{xy} \sim \sum_{fi} \left(1-\Delta m\right) g_{fi}
\left\vert \langle f \vert \hat{v}_x \vert i \rangle \right\vert^2 \ , \\
\chi_L &=&  \chi_{xx}+i\chi_{xy} \sim \sum_{fi} \left(1+\Delta m\right) g_{fi}
\left\vert \langle f \vert \hat{v}_x \vert i \rangle \right\vert^2 \ . 
\en
In these expressions, only $\Delta m = -1$ ($\Delta m = +1$) contributes to $\chi_R$ ($\chi_L$).
Each factor is a Kronecker delta, i.e., $(1 \pm \Delta m) = 2\delta_{m_f=m_i \pm 1}$.
The following expressions result for integration in the band model expressed using the 
transitions between occupied (lower band) and unoccupied (higher band) states:
\begin{widetext}
\bn
\label{chiRL}
\chi_{R} &=&
\frac{2ne^2}{\epsilon_0\hbar(\omega+i\gamma)} \sum_{i}^{o}\sum_{f}^{u}
\frac{w_i-w_{f}}{\omega_{if}} \left\vert\langle f\vert \hat{v}_x\vert i\rangle \right\vert^2
\left\{ \frac{\delta_{m_f=m_i-1}}{\omega+i\gamma+\omega_{if}}
+ \frac{\delta_{m_f=m_i+1}}{\omega+i\gamma-\omega_{if}} \right\} \ , \\
\chi_{L} &=&
\frac{2ne^2}{\epsilon_0\hbar(\omega+i\gamma)} \sum_{i}^{o}\sum_{f}^{u}
\frac{w_i-w_{f}}{\omega_{if}} \left\vert\langle f\vert \hat{v}_x\vert i\rangle \right\vert^2
\left\{ \frac{\delta_{m_f=m_i+1}}{\omega+i\gamma+\omega_{if}}
+ \frac{\delta_{m_f=m_i-1}}{\omega+i\gamma-\omega_{if}} \right\} \ .
\en
\end{widetext}
The only difference between these is the swapping of the Kronecker deltas. Then the two cases can 
be written in terms of a single expression, replacing the $\pm 1$ in the Kronecker deltas with the
helicity index:  
\bn 
\label{chinu1}
\chi_{\nu} &=&  \frac{2ne^2}{\epsilon_0\hbar(\omega+i\gamma)} \sum_{i}^{o}\sum_{f}^{u}
\frac{w_i-w_{f}}{\omega_{if}} \left\vert\langle f\vert \hat{v}_x\vert i\rangle \right\vert^2
\nonumber \\
&& \times 
\left\{ \frac{\delta_{m_f=m_i+\nu}}{\omega+i\gamma+\omega_{if}}
+ \frac{\delta_{m_f=m_i-\nu}}{\omega+i\gamma-\omega_{if}} \right\} \ .
\en
To proceed further, it is necessary to evaluate the sums.  This can be facilitated by converting
them to integrals over the allowed transitions, which depends slightly on the dimensionality of the
bands under consideration.

\subsection{Interband transitions between three-dimensional bands}
The band structure of interest could be effectively isotropic and three-dimensional, 
say, for the case of some semiconductors near the $\Gamma$ point (${\bf k}=0$).
Therefore it is interesting to consider the IBT contribution for this model, before
doing a similar analysis of the reduced one-dimensional band model for metals.

Converting from a sum to an integral with $\sum_{\bf k}\rightarrow \frac{V}{(2\pi)^3}\int d{\bf k}$, 
using the assumed form for the matrix elements, and then changing to $s=\sqrt{\hbar/2m^*}\, k$ 
as the variable of integration, the interband susceptibility can be written
\be
\chi_{\nu} = Q\, T_{\nu}(\omega)\ ,
\ee
where $Q$ contains all the constant normalization factors, and the interband transition integral
$T_{\nu}(\omega)$ contains all of the frequency and temperature dependence:
\bn
\label{Q3D}
Q &=&  
\frac{2ne^2 |M|^2\hbar}{m_o^2 \epsilon_0} \frac{V}{(2\pi)^3} \frac{4\pi}{3}
\left(\frac{2m^*}{\hbar}\right)^{5/2}\ , \\
\label{Tnu3D}
T_{\nu} &=& \frac{1}{\omega+i\gamma}
 \sum_{m_i}\sum_{m_f} \int_{0}^{s_F} ds\ \frac{w_i-w_{f}}{\omega_{if}}\, s^4 \nonumber \\
&& \times \left\{ \frac{\delta_{m_f=m_i+\nu}}{\omega+i\gamma+\omega_{if}}
+ \frac{\delta_{m_f=m_i-\nu}}{\omega+i\gamma-\omega_{if}} \right\} \ ,
\en
To arrive at this, $\int d\Omega\, k_x^2 = \frac{4\pi}{3} k^2$ was used for the angular part of the 
integration.  The upper limit is a Fermi wave vector $s_F=\sqrt{\hbar/2m^*}\, k_F$ needed to 
sum over all the states of the occupied initial band. We can let $w_i=1$ for the lower band, but keep 
the the temperature-dependent occupation probability $w_f>0$ for the upper band.  In this way, any 
thermal effects due to an initial population in the final band will be included.

The first integral in (\ref{Tnu3D}) uses $\Delta m = +\nu$ and the second uses $\Delta m = -\nu$.
These choices enter in the expression for $\omega_{if}(\Delta m)$.  In terms of the scaled wave 
vector $s$ or a related excitation variable $x=\omega_g+s^2$, one has
\bn
\omega_{if}(+\nu) &=& -\omega_g-s^2+\nu\frac{\omega_B}{2} = -x+\zeta_{\nu} \ , \\
\omega_{if}(-\nu) &=& -\omega_g-s^2-\nu\frac{\omega_B}{2} = -x-\zeta_{\nu} \ .
\en
The variable $x$ is the excitation energy above the lower band, and $\zeta_{\nu}=\tfrac{1}{2}\nu\omega_B$
is the polarization-dependent Zeeman splitting. Then the denominators in (\ref{Tnu3D}) in the two
terms are $(\omega+i\gamma+\zeta_{\nu}-x)$ and $(\omega+i\gamma+\zeta_{\nu}+x)$, respectively.
This suggests introducing notation for a Zeeman-shifted complex optical frequency, for the two 
circular polarizations:
\be
\label{omnu}
\omega_{\nu} \equiv \omega+i \gamma +\zeta_{\nu} = \omega+i \gamma +\tfrac{1}{2}\nu \omega_B \ .
\ee
Thus, most of the polarization-dependent effects will be carried by this shifted frequency.

To include the Fermi occupation factor $w_i-w_f$, the sum over $m_i$ in (\ref{Tnu3D}) can be done first,
holding $m_f$ fixed.  Then both terms will have the same occupation factor, taken to be $1-w_f$. The 
first term in (\ref{Tnu3D}) uses only $m_i=m_f-\nu$, the second uses only $m_i=m_f+\nu$, assuming those 
states exist in the lower band.  The final state energy, however, depends on the effective mass $m^*_e$ 
in the upper band, whereas the transition energy depends on the reduced mass $m^*$.  If these are 
nearly the same, i.e., $m^*_e \approx m^*_h \approx m^*$, then the final state energy measured relative
to the top of the lower band can be taken as
\be
E_f\approx \hbar\left( \omega_g+s^2-\tfrac{1}{2}m_f \omega_B\right) 
= \hbar \left(x-\tfrac{1}{2}m_f \omega_B\right) .
\ee
This will lead to an occupation factor $w_i-w_f$ for each term determined from the Fermi energy $E_F$,
\be
g_{m_f}(x) \approx 1-F(E_f-E_F)=1-F(x;m_f).
\ee

Inouye et al.\cite{Inouye++98} and also Scaffardi and Tocho\cite{Scaffardi06} have discussed 
bound electron response in a 1D band model without ${\bf B}$, applying an expression which is
an integral over the excitation variable $x$.  We can transform the integral $I_{\nu}$ into an 
integration over $x$ to compare with those results.  In the simplest case, when the lower
band has a higher angular momentum $l_i$ than the upper band value, $l_f$, then both states
$m_i=m_f\pm\nu$ are available for all the allowed $m_f$.  In this case, Eq.\ (\ref{Tnu3D}) 
becomes
\be
\label{Tnux3D}
T_{\nu} = \frac{\omega_{2\nu}}{\omega+i\gamma} \sum_{m_f} \int_{\omega_g}^{x_F} dx\ 
\frac{ g_{m_f}(x)\, x \left(x-\omega_g\right)^{3/2}}
{(x^2-\tfrac{1}{4}\omega_{B}^2)(x^2-\omega_{\nu}^2)}\ .
\ee
The function $g_{m_f}=w_i-w_f$ includes the magnetic field effects on the final state occupation 
probabilities, which are slightly different for each $m_f$ level. The upper limit is determined by 
the scaled wave vector at the Fermi level, $x_F=\omega_g+s^2_F$.  Although this is the a 3D band 
model, the similarity to the corrected expression from Scaffardi and Tocho\cite{Scaffardi06} is 
clear. When reduced to one dimension (use $\sqrt{x-\omega_g}$ in place of 
$\left(x-\omega_g\right)^{3/2}$, together with a different constant factor $Q$ out front), it 
recovers the expression from Ref. \onlinecite{Scaffardi06} for $\omega_B\rightarrow 0$.
 
The main effect of the DC magnetic field is to shift the optical frequency oppositely for the two 
circular polarizations. The second important effect is the modification of the $m_f$ occupations
with $B$. There appears another effect due to the factor $x^2-\tfrac{1}{4}\omega_{B}^2$ 
in the denominator, however, that is relatively small, quadratic in the field, and not dependent 
on the polarization.  The dependence on $2\nu$ in the numerator is interesting.  

In the limit where the initial states are fully occupied and the final states are fully unoccupied 
(all $w_f=0$), these integrals can be carried out analytically, see Appendix \ref{APP3D}. In the
further approximation where the upper limit of integration is set to $s_F\rightarrow \infty$, the 
result for this complex integral is:
\bn
T_{\nu} &=&
  \frac{\pi (l_i+l_f) }{2(\omega+i\gamma)^2}
\left\{i(\omega_{\nu}-\omega_g)^{3/2}+(\omega_{\nu}+\omega_g)^{3/2} \right.  \nonumber \\
&& \left.  -(\omega_g-\zeta_{\nu})^{3/2} - (\omega_g+\zeta_{\nu})^{3/2} \right\} .
\en
The lower and upper band angular momenta combine to give the multiplicity of transitions, $g_m=l_i+l_f$.
This result corresponds to the situation of intrinsic semiconductor particles where the Fermi level is near
or below the middle of the gap.  This limit takes out any temperature dependence.  However, that should be 
a small effect only from the rounding of the occupation of levels near the Fermi energy. Assuming that the 
gap frequency $\omega_g$ is large compared to the Zeeman splitting $\omega_B$, temperature effects 
on the FR could be small.  Furthermore, the limit of zero damping is simple to read out from this 
expression.  However, generally, we can calculate all results more precisely using the full theory result
in equation (\ref{Tnux3D}).

\subsection{A 1D band model}
The band structure in real materials can be very complicated.  The isotropic 3D model just 
discussed is an idealization for real solids.  In the band structure of gold\cite{CS71} in the 
L direction at the Fermi surface, the important IBTs occur from the d valence band to the 
sp-conduction band, where the two bands are separated by a gap of $E_g\approx 2$ eV. Because 
this happens along a particular direction, a 1D band model is useful.  Inouye \textit{et al.}
\cite{Inouye++98} used a 1D band model to get the absorption contribution in gold due to 
the interband transitions for bound electrons.  They were able to fit the absorption very well for
photon energies above 2 eV up into the ultraviolet, well above the small particle plasmon
resonance frequency.  Without interband transitions taken into account, it is impossible to get
such an accurate description. 

The main difference from the 3D problem, is that the integration may start out in 3D,
but needs to be reduced to an effective integration along only the active direction of the 
band, which is taken as the $k_x$-axis.  The details of how this is done are not so important.
One way is to convert the sum in (\ref{chinu1}) into integration in a box along Cartesian axes, using
that to define an effective Fermi wave vector (slightly different from the standard definition) by
\be
N= \frac{2V}{\left(2\pi\right)^3}  \int_{-k_F}^{k_F}
 \int_{-k_F}^{k_F} \int_{-k_F}^{k_F} dk_x\, dk_y\, dk_z = \frac{2Vk_F^3}{\pi^3} \ .
\ee
Then an effective Fermi wave vector is defined here as $k_F=\pi(n/2)^{1/3}$, which differs from 
the usual expression $k_F=(3\pi^2 n)^{1/3}$ by the factor $(\pi/6)^{1/3}\approx 0.8$, of no
real importance.  Now in the sum over wave vectors in (\ref{chinu1}), the transverse (inactive) 
coordinates $y,z$ are integrated out, leaving only the integration over the band coordinate:
\be
\sum_{{\bf k}_i} \rightarrow \frac{2V}{(2\pi)^3} (2k_F)^2 \int_{-k_F}^{k_F} dk_x
=\frac{Vk_F^2}{\pi^3} \int_{-k_F}^{k_F} dk_x \ .
\ee
Normalized this way, this final integration will correctly give a dimensionless result for $\chi_{\nu}$.
The conversion from sum to integration is carried out by including the factor $V k_F^2/\pi^3$ rather 
than the more usual factor $V/(2\pi)^3$ used for 3D.  Then for 1D, the interband susceptibility can also 
be written $\chi_{\nu}=Q T_{\nu}(\omega)$, the latter being an integration over $s=\sqrt{\hbar/2m^*}\, k_x$,
and containing the frequency and temperature dependence,
\bn
\label{Q1D}
Q &=&
 \frac{2ne^2|M|^2\hbar}{m_o^2\epsilon_0}\frac{2 V k_F^2}{\pi^3} 
\left(\!\frac{2\tilde{m}}{\hbar}\!\right)^{3/2} \ , \\
\label{Tnu1D}
 T_{\nu} &=& \frac{1}{\omega+i\gamma} \sum_{m_i}\sum_{m_f} \int_{0}^{s_F} ds ~
\frac{w_i-w_f}{\omega_{if}} ~ s^2 \nonumber \\
&& \times \left\{ \frac{\delta_{m_{f}=m_{i}+\nu}}{\omega+i\gamma+\omega_{if}}
 + \frac{\delta_{m_{f}=m_{i}-\nu}}{\omega+i\gamma-\omega_{if}} \right\} \ .
\en
This is very similar to the expression (\ref{Tnu3D}) for 3D. The only difference is the factor
$s^2$ for 1D in place of $s^4$ in 3D.  

The expression for $T_{\nu}$ can be written with the excitation variable $x=\omega_g+s^2$; 
In the most general case, both states $m_i=m_f\pm \nu$ do not exist for all choices of $m_f$, 
and the two terms in the integrand cannot be combined in a simple form. One must then evaluate 
(\ref{Tnu1D}) by
\bn
 T_{\nu} &=& \frac{-1}{\omega+i\gamma} \sum_{m_f} \int_{\omega_g}^{x_F} dx ~ g_{m_f}(x)\,
\frac{\sqrt{x-\omega_g}}{x-\zeta_{\nu}} ~ \times \nonumber \\
 && \times \frac{1}{2}\sum_{m_i} \left\{ \frac{\delta_{m_{i}=m_{f}-\nu}}{\omega_{\nu}-x}
 + \frac{\delta_{m_{i}=m_{f}+\nu}}{\omega_{\nu}+x} \right\} \ .
\en
When the angular momentum $l_f$ of the upper band is less than that of the lower band, $l_i$,
both the choices $m_i=m_f\pm \nu$ always exist for any $m_f$, and this can be written instead as  
\be
T_{\nu}=\frac{\omega_{2\nu}}{\omega+i\gamma} \sum_{m_f} \int_{\omega_g}^{x_F} dx \  
\frac{g_{m_f}(x)\, x \sqrt{x-\omega_g}}
{(x^2-\tfrac{1}{4}\omega_B^2)(x^2-\omega_{\nu}^2)} \ .
\ee
As mentioned earlier, the only change from the 3D expression is that the power is now $\tfrac{1}{2}$ 
instead of $\tfrac{3}{2}$ in the numerator.   That is a complex integrand. If the real and imaginary parts
are needed,  it can be written in terms of real quantities, better for comparison with the $B=0$ equations:
\bn
T_{\nu} &=& \frac{\omega_{2\nu}}{\omega+i\gamma} \sum_{m_f} \int_{\omega_g}^{x_F} dx~ g_{m_f}(x) ~ 
\frac{x \sqrt{x-\omega_g}}{x^2-\tfrac{1}{4}\omega_B^2} \nonumber \\
&& 
\label{Tnux1D}
\times \frac{ \left(x^2+\gamma^2-\omega_F^2 \right) +2i\gamma \omega_F}
{\left(x^2+\gamma^2-\omega_F^2 \right)^2 + 4\gamma^2 \omega_F^2} \ .
\en
Here a notation for a real shifted Faraday frequency is used,
\be
\omega_{F} \equiv \omega+\tfrac{1}{2}\nu\omega_B \ ,
\ee
that is dependent on the polarization index $\nu$.  In the application
of this to gold nanoparticles, we assume transitions from the lower d-band to the upper p-band,
hence there are values $m_f=-1,0,1$ to be summed over.

The last result also has an approximate expression that applies if the Fermi energy falls well within
the band gap.  Then, the Fermi occupation factors can be approximated with $w_i-w_f=1$, removing all
the temperature dependence.  The integral for $T_{\nu}$ can be done exactly in this case, see the Appendix.
There results
\bn
\label{Tnu1D-approx}
T_{\nu} &=&
\frac{\pi (l_i+l_f)}{2(\omega+i\gamma)^2}
\left\{ i \sqrt{\omega_{\nu}-\omega_g} - \sqrt{\omega_{\nu}+\omega_g} \right.  \nonumber \\
&& \left. + \sqrt{\omega_g-\zeta_{\nu}} + \sqrt{\omega_g+\zeta_{\nu}} \right\} .
\en
Obviously the expression is similar to that for the 3D band model, but now the powers are 
$\tfrac{1}{2}$ instead of $\tfrac{3}{2}$.  A comparison of the full integral (\ref{Tnux1D})
and this approximation (\ref{Tnu1D-approx}), as functions of the photon energy, can be seen in the
Appendix in Figure \ref{eps-ibt-eV}.

\subsection{Interband parameters for gold}
\begin{figure}
\begin{center}
\includegraphics[width=\smallfigwidth,angle=-90]{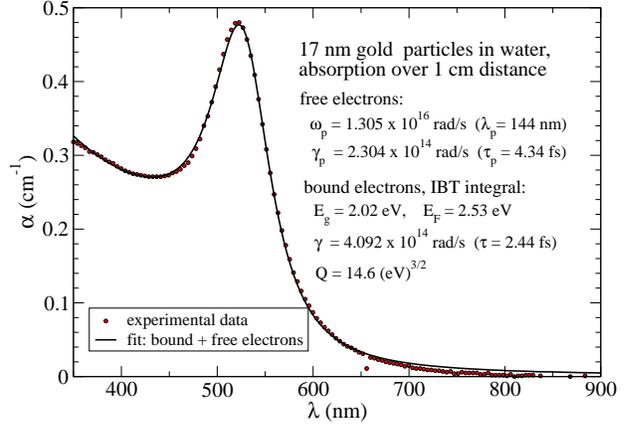}
\end{center}
\caption{\label{gold-abs-ibt} (Color online) Fitting of the absorption of 17 nm diameter gold particles in
water solution, according to the 1D band model for the interband dielectric response. Fit was made via
a Monte Carlo search, allowing both the bound electron and free electron parameters to be varied.
Their final adjusted values, for $T=300$ K, are indicated on the Figure. The fitted gold volume
fraction is $f_s=5.95 \times 10^{-7}$.}  
\end{figure}
\begin{figure}
\begin{center}
\includegraphics[width=\smallfigwidth,angle=-90]{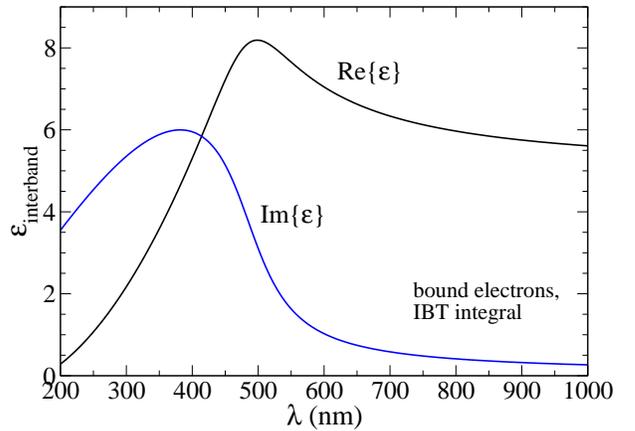}
\end{center}
\caption{\label{eps-ibt} (Color online) The IBT contribution to the permittivity, from the
1D band model, Eq.\ (\ref{Tnux1D}), using the parameters of Figure \ref{gold-abs-ibt}. 
In this model both the real and imaginary parts remain positive for all frequencies.}
\end{figure}

Following Inouye {\it et al.}\cite{Inouye++98} and also Scaffardi and Tocho\cite{Scaffardi06}, 
the 1D band model was applied here for the dielectric response of gold nanoparticles.  As for
the Drude model, we can fit the 1D interband permittivity, Eq.\ (\ref{Tnux1D}) for $B=0$ to the
absorption of a solution of 17 nm diameter gold particles in water.  The absorption data
from 350 nm -- 900 nm were fit to Eq.\ (\ref{Tnux1D}) (using also the MG theory presented earlier)
while allowing the gap energy $\hbar\omega_g$, Fermi energy $E_F$, bound electron damping 
$\gamma$, and normalization constant $Q$ to be varied. We also allow the gold volume fraction
$f_s$ and free electron plasma frequency $\omega_p$ and damping $\gamma_p$ to be varied. 

This is a multi-parameter
search, which was carried out via a Metropolis Monte Carlo algorithm. For the effective energy function 
to be minimized, it was found practical to use the sum of the absolute differences between the
experimental data $\alpha_i$ at each frequency and the theoretical expression $\alpha_{\rm th}$, i.e., 
$\sum_i \vert \alpha_i(\omega_i)-\alpha_{\rm th}(\omega_i)\vert$, instead of the squares (this
gives a more uniform weighting but looser fit to the points).  A reasonable fit and the associated 
parameters are shown in Figure \ref{gold-abs-ibt}. The gap energy $E_g=2.02$ eV, Fermi energy 
$E_F=2.53$ eV and plasma frequency ($\hbar\omega_p=8.59$ eV) determined in this fit are consistent with 
the values used  in Ref. \onlinecite{Scaffardi06}.  The dampings, $\hbar\gamma_p=0.152$ eV 
($\gamma_p^{-1}=4.33$ fs) and $\hbar \gamma=0.269$ eV ($\gamma^{-1}=2.45$ fs), are somewhat different 
from those for bulk gold, and the resulting real and imaginary parts of dielectric function \ew
are different from those for bulk gold.\cite{JC72} However, $\gamma_p$ is consistent with the prediction 
from Eq.\ (\ref{gp}), as expected due to extra surface scattering and other factors for the nanoscale particles.  
If this surface scattering effect is taken out, the model reproduces the real and imaginary parts of
\ew for bulk gold that are completely consistent with those found by Johnson and Christy.\cite{JC72}

A value of gold volume fraction $f_s=5.95 \times 10^{-7}$ was needed in the fit. It is about half
the value estimated by the techniques in Ref.\ \onlinecite{Liu+07}, see Sec.\ \ref{synthesis}, 
showing the difficulty in estimation of $f_s$ in the lab.  We might note that this fit is not 
strongly constrained; it was determined only by the absorption data; other 
values cannot be strongly ruled out.  Unlike the Drude model presented earlier, this model is very 
good at fitting the ultraviolet end of the dielectric properties, and also fits the infrared end better.  
Thus, we expect it should give more reliable predictions for the Faraday rotation properties.

\section{Faraday rotation due to gold nanoparticles}
Here we apply the 1D band model to the interband transitions, and compare the theory
to experiments for gold NPs.  Based on the fitting of parameters for 17 nm  diameter 
gold particles in the previous section, the results for scaled complex Faraday rotation angle 
$\Psi=\Upsilon+iZ = (\varphi+i{\cal X})/(Bzf_s)$ can be estimated using Eq.\ (\ref{psi}) together 
with the MG theory for the response of a dilute solution, Eq.\ (\ref{epseff}).  
The real and imaginary parts are the  Verdet and ellipticity factors, respectively,
scaled by volume fraction $f_s$ of gold in the solution.

\begin{center}
\begin{figure}
\includegraphics[width=\smallfigwidth,angle=-90]{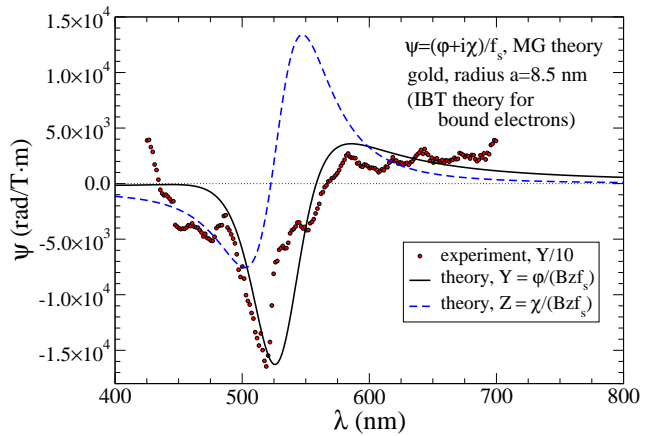}
\caption{\label{psi-gold17-ibt} (Color online)  (a) The real and imaginary parts of the scaled 
complex Faraday rotation $\Psi=(\varphi+ i {\cal X})/(B z f_s)$, from experiment (points), and calculated 
including IBTs for bound electrons in the 1D band model for gold particles of 17 nm diameter (lines).  
The real part (solid curve) is the Verdet factor per volume fraction, $\Upsilon$.  The experimental 
data for $\Upsilon$ has been scaled by a factor of $1/10$, which brings its peak to the same size as in 
the IBT theory. The imaginary part is the ellipticity factor scaled by volume fraction (not measured).}
\end{figure}
\end{center}

The theory results for 17 nm diameter gold NPs are compared with experimental data for
the FR spectrum in Fig.\ \ref{psi-gold17-ibt}.  The experimental data for
$\Upsilon=\upsilon/f_s$ has been scaled down by a factor of $1/10$ so that its negative FR
peak is similar in magnitude to the theory.  Although some details are not 
identical between theory and experiment, the 
general trend in $\Upsilon$ as a function of wavelength is similar.  Both show a strong
negative peak due to the plasmon near 520 -- 530 nm, and a zero crossing in $\Upsilon$ near
550 nm.  This peak is about ten times stronger when using the IBT theory than the simpler
Drude approach presented in Fig.\ \ref{psi-gold17-drude}, but the negative FR peak in the
experimental data is still much stronger than in both theories.  The experimental result 
for $\Upsilon$, however, is obtained by dividing raw FR data by the estimated gold volume fraction
(using techniques of Ref.\ \onlinecite{Liu+07}), $f_s=1.23\times 10^{-6}$.  Any error in the 
volume fraction will modify the experimental value of $\Upsilon$. The volume fraction  divides
out in the theory for $\Upsilon$.   Various multiple scattering and backscattering effects\cite{Gas+13}
and other similar aggregation effects\cite{Pakdel+12} not included in this theory could explain 
this large discrepancy.

Above 560 nm both theory and experiment indicate a positive Faraday rotation angle.
According to the theory, there is also a strong positive peak in ellipticity expected around  
a wavelength slightly larger than that due to the plasmon (around 550 nm). Notably, both the
rotation and ellipticity tend towards zero at short wavelengths, removing the 
artifact present in the classical Drude approach.

\section{Discussion and Conclusions}
The interband electronic transitions are known to have a considerable effect on dielectric
properties of gold and other metals.  Although it is popular to consider only a
simple Drude model for quasi-free electrons, it is shown here to be inadequate for
describing, say, the absorption in gold nanoparticles.  A fit for the dielectric parameters 
based on a Drude model, combining free electron and bound electron contributions, was given 
in Fig.\ \ref{gold-abs-drude}, for a dilute solution of gold NPs.  The absorption peak near 
520 nm, associated with the excitation of a plasmon resonance in the NPs, can be fit rather 
well.  In the violet and ultraviolet, however, it is impossible to get a good match 
between the theory and experimental data for gold particles of average diameter of 17 nm.  
This approach leads (incorrectly) to negative values of the real part of permittivity at shorter
wavelengths (below the plasmon resonance), which is why the fit fails in that wavelength region.

For the optical effects such as absorption or Faraday rotation, it is clear that a correct 
description of $\epsilon(\omega)$ for a macroscopic sample completely determines the outcome.  
If the absorption cannot be fitted properly, then it is hard to see how the Drude approach 
could explain Faraday rotation very well.  Instead, we include the
interband transitions for the bound electrons, based on the quantum formulation due to
Boswarva \textit{et al.}\cite{BHL62} and also Adler.\cite{Adler62}.  However, the calculation
done here is specifically for nanoparticles in the presence of a DC magnetic field $B$. 
Bulk calculations\cite{BHL62,Halpern+64} have assumed Landau levels due to $B$ for an infinite
sample, but the Landau radius $r_0=\sqrt{\tfrac{2\hbar}{eB}}$ for weak $B$ will be larger
than the typical NP size.  Hence, the Landau levels have no physical sense for weak magnetic field; they
are not the quantum energy eigenstates.  The states are modified due to the geometric confinement
of a limited size nanoparticle.  The shift in energy states used here is a Zeeman shift due
to the DC magnetic field for electrons in bands of specified angular momentum.  

We are applying this calculation at a magnetic field strength such that the Landau radius is 
only about twice the NP radius.  This more intermediate field is needed to insure adequate
signal to noise ratio in the FR data.  This should be within the range of the quantum theory.
However, at a magnetic field strength about four times larger, the Landau radius will match
the NP radius, and crossover to a more bulk-like behavior could be expected.

We have sketched out the  details of the IBT integral, especially for the 1D band model.
While it was required to fit the amplitude factor $Q$ and other dielectric parameters,
this results in a much better fit between theory and experiment for the absorption,
Fig.\ \ref{gold-abs-ibt}.  The fitting leads to an interband permittivity whose real
and imaginary parts remain positive for $\lambda>200$ nm, Fig.\ \ref{eps-ibt}.  
Thus we expect this approach to be good when applied to the Faraday rotation
properties of a dilute solution of NPs.  

When applied to a solution of 17 nm diameter gold NPs, one finds a strong negative peak 
in the Verdet factor, as expected, near the plasmon wavelength of about 525 nm.  This negative
peak is certainly due to extra Faraday rotation that is associated with the surface
plasmon mode in nanoparticles. This is at least partially confirmed by experiments at 
a small volume fraction of gold.  The theory also predicts a positive peak in ellipticity at 
a wavelength somewhat larger than the plasmon wavelength (nearly 550 nm for 17 nm gold NPs).  
However, at present, the experimental results for scaled Verdet factor 
$\Upsilon=\upsilon/f_s$ seem to be about an order of magnitude stronger than that indicated 
by this theory.  Some of this discrepancy could be due to uncertainty in the true value
of gold volume fraction.  There may also be other processes present in the NP data, 
such as aggregation\cite{Pakdel+12} and backscattering effects,\cite{Gas+13}
 that are not included in the model.  At present we do not have comparable experimental 
results on the ellipticity factor.  It will be important to clarify this discrepancy in 
the size of the magneto-optical responses.
 
\section*{Acknowledgments}
GMW is grateful for the hospitality of Universidade Federal de Vi\c cosa in Vi\c cosa, 
Minas Gerais, Brazil, and Universidade Federal de Santa Catarina in Florian\'opolis, Brazil, 
where parts of this work were carried out, and financial support of FAPEMIG grant BPV-00046-11.
We acknowledge the support of NSF through grants NSF-930673 and NSF-1128570, and the 
Terry Johnson Cancer Center at KSU for funding the construction of the pulsed magnet.

\appendix

\section{Evaluation of integrals over scaled wave vector}
\label{APP3D}
This analysis applies when the Fermi level is well within the band gap, and then
the upper band is assumed to be unoccupied.  This removes the temperature dependence from
the model.

\subsection{3D band model}
The integral for the 3D band model, equation (\ref{Tnu3D}), is best evaluated using 
$s=\sqrt{\hbar/2m^*}\, k$ as the variable of integration. We assume $w_i=1, \quad w_f=0$,
and initially include the damping parameter $\gamma$. For the sum over $m_f$, the Kronecker deltas 
select $m_f=m_i\pm \nu$ for each integrand.  For the sum over $m_i$, between bands with defined
orbital angular momentum $l_i$ and $l_f$, there are multiple equivalent transitions, all with
the same thermodynamic weight. This leads to a multiplicity
\be
g_m=\sum_{m_i} 1 = {\rm min}(2l_i+1,2l_f+1) = l_i+l_f \ .
\ee
Essentially, this constant factor replaces the temperature dependence that would have been
included by the occupation function, $g_{m_f}(x)$ in the full theory expression.  
Then in this approximation, the interband susceptibility is $\chi_{\nu}=Q T_{\nu}(\omega)$, 
with $Q$ defined earlier in (\ref{Q3D}), and the transition integral expressed
\bn
T_{\nu} =
\frac{g_m}{\omega+i\gamma} \,  ( K_1 + K_2 ) \ .
\en
$T_{\nu}$ has been split into two similar integrals,
\bn
K_1 &=& \int_{0}^{s_F} \frac{+ds\, s^4}{(s^2+\omega_g-\zeta_{\nu})(s^2+\omega_{g}-\omega_{\nu})} \ ,
\\
K_2 &=& \int_{0}^{s_F} \frac{-ds\, s^4}{(s^2+\omega_g+\zeta_{\nu})(s^2+\omega_{g}+\omega_{\nu})} \ . 
\en
The shifted frequency $\omega_{\nu}$ is defined in (\ref{omnu}) and $\zeta_{\nu}=\tfrac{1}{2}\nu\omega_B$.
Consider $K_1$ using the partial fraction expansion,
\be
K_1 = \int_{0}^{s_F} \!\!\! \frac{ds\, s^4}{\omega_{\nu}-\zeta_{\nu}} \left[ 
\frac{1}{s^2+\omega_g-\omega_{\nu}} -\frac{1}{s^2+\omega_g-\zeta_{\nu}} \right] \ .
\ee
The integral $K_2$ is the negative of this with $\zeta_{\nu}$ and $\omega_{\nu}$ reversed in sign:
\be
K_2 = \int_{0}^{s_F} \!\!\! \frac{ds\, s^4}{\omega_{\nu}-\zeta_{\nu}} \left[ 
\frac{1}{s^2+\omega_g+\omega_{\nu}} -\frac{1}{s^2+\omega_g+\zeta_{\nu}} \right] \ .
\ee
Note that $\omega_{\nu}-\zeta_{\nu}=\omega+i\gamma$.
These integrals can all be found from the indefinite integral,
\be
f(s)=\int \frac{ds\, s^4}{s^2+a^2} = \frac{1}{3}s^3-a^2s+a^3 \tan^{-1}\left(\frac{s}{a}\right) \ ,
\ee
where the parameter $a$ is complex for the four different cases where this is used. Thus the function 
$f(s)$ is defined with the analytic continuation of the inverse tangent to complex arguments.  Applying 
this gives
\bn
K_1 &=& \frac{1}{\omega+i\gamma} \left\{ 
(\omega_g-\omega_{\nu})^{3/2}
\tan^{-1}\left(\frac{s}{\sqrt{\omega_g-\omega_{\nu}}}\right) \right. \nonumber \\
&-&  \left. (\omega_g-\zeta_{\nu})^{3/2}
\tan^{-1}\left(\frac{s}{\sqrt{\omega_g-\zeta_{\nu}}}\right) \right\} +s \ , \\
K_2 &=& \frac{1}{\omega+i\gamma} \left\{ 
(\omega_g+\omega_{\nu})^{3/2}
\tan^{-1}\left(\frac{s}{\sqrt{\omega_g+\omega_{\nu}}}\right) \right. \nonumber \\
&-&  \left. (\omega_g+\zeta_{\nu})^{3/2}
\tan^{-1}\left(\frac{s}{\sqrt{\omega_g+\zeta_{\nu}}}\right) \right\} -s \ .
\en
It is interesting that once these are summed to produce $T_{\nu}=\frac{g_m}{\omega+i\gamma}(K_1+K_2)$, all linear
and cubic terms in $s$ cancel out, leaving only the inverse tangents.  These are evaluated at
the upper limit $s_F$.  Due to these cancellations, the upper limit can be let to go to infinity
as a reasonable approximation (also, the choice of a finite $s_F$ may be difficult in any case).
With $s$ real, the limit of the complex inverse tangent can be shown to have the following 
dependence on the parameter $a$:
\bn
\lim_{s\rightarrow \infty} \tan^{-1} \frac{s}{a} &=& \sgn\left(\text{Re}\{a\}\right)\frac{\pi}{2}\ , \\
\lim_{s\rightarrow \infty} \tan^{-1} \frac{is}{a} &=& \sgn\left(\text{Im}\{a\}\right)\frac{\pi}{2}\ .
\en
The latter form is useful for the first $\tan^{-1}$ in $K_1$, changing $\sqrt{\omega_g-\omega_{\nu}}$
to $i\sqrt{\omega_{\nu}-\omega_g}$, which is more convenient when assuming above the gap excitation.
Then the resulting interband integral for the 3D band model under these approximations is 
\bn
T_{\nu} &=&
\frac{\pi g_m}{2(\omega+i\gamma)^2}
\left\{i (\omega_{\nu}-\omega_g)^{3/2}+(\omega_{\nu}+\omega_g)^{3/2} \right.  \nonumber \\
&& \left. - (\omega_g-\zeta_{\nu})^{3/2} - (\omega_g+\zeta_{\nu})^{3/2} \right\} \ .
\en
These complex square roots are defined by the root with the positive real part, i.e., the root in the 
first or fourth quadrant.  It is interesting to notice that the photon frequency does not appear inside 
the last factors involving the Zeeman shift $\zeta_{\nu}$.  Also, in the limit of zero damping $\gamma\rightarrow 0$, 
the very first term is the entire imaginary part.  For below the gap excitation ($\omega_{\nu}<\omega_g$) 
the imaginary part becomes zero in the absence of damping.  For typical material parameters, this approximation
gives results very close to the full theoretical result from Eq.\ (\ref{Tnux3D}).  This is mostly due to the
high dimensionality; in the 1D band model this approximation is farther from the full theory expression.

\subsection{1D band model}
\begin{figure}
\begin{center}
\includegraphics[width=\smallfigwidth,angle=-90]{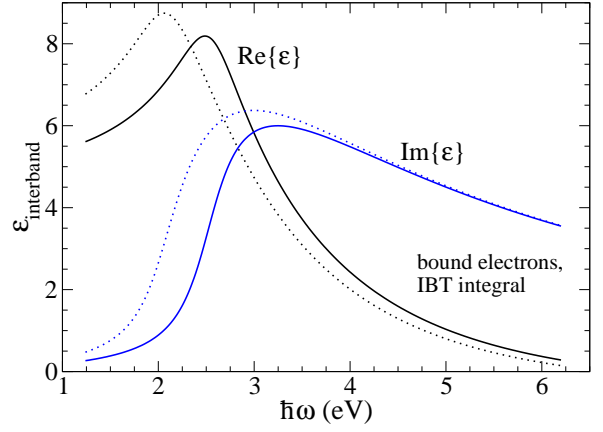}
\end{center}
\caption{\label{eps-ibt-eV} (Color online) The IBT contribution to the permittivity for the
1D band model, as functions of photon energy in eV, using the parameters of Figure \ref{gold-abs-ibt}.
The solid curves apply the full theory, Eq.\ (\ref{Tnux1D}), including the Fermi occupation factors,
for temperature 300 K.  Note the peak in Re$\{\epsilon\}$ at the Fermi energy (2.52 eV) in the
full theory.  The dotted curves show the result of assuming a totally unoccupied upper band, $w_f=0$, 
obtaining \ew from Eq.\ (\ref{Tnu1D-semi}).}
\end{figure}

The analysis is nearly the same, with interband susceptibility expressed via $\chi_{\nu}=QT_{\nu}(\omega)$.  
One has the same expression for the multiplicity, $g_m=l_i+l_f$.  The weighting factor $Q$ is now given in 
equation (\ref{Q1D}).  The expression for the transition integral can still be written
\be
T_{\nu} = \frac{g_m}{\omega+i\gamma} (K_1+K_2)\ ,
\ee
except that in this case, the parts have the power $s^2$ in place of $s^4$ in their numerators:
\bn
K_1 &=& \int_{0}^{s_F} \frac{+ds\, s^2}{(s^2+\omega_g-\zeta_{\nu})(s^2+\omega_{g}-\omega_{\nu})} \ ,
\\
K_2 &=& \int_{0}^{s_F} \frac{-ds\, s^2}{(s^2+\omega_g+\zeta_{\nu})(s^2+\omega_{g}+\omega_{\nu})} \ . 
\en
The partial fraction expansions are
\be
K_1 = \int_{0}^{s_F} \!\!\! \frac{ds\ s^2}{\omega_{\nu}-\zeta_{\nu}} \left[
\frac{1}{s^2+\omega_g-\omega_{\nu}} -\frac{1}{s^2+\omega_g-\zeta_{\nu}} \right] \ ,
\ee
and for $K_2$, reverse the overall sign and the signs on $\omega_{\nu}$ and $\zeta_{\nu}$,
\be
K_2 = \int_{0}^{s_F} \!\!\! \frac{ds\ s^2}{\omega_{\nu}-\zeta_{\nu}} \left[
\frac{1}{s^2+\omega_g+\omega_{\nu}} -\frac{1}{s^2+\omega_g+\zeta_{\nu}} \right] \ .
\ee
The basic integral needed is
\be
f(s)=\int \frac{ds\ s^2}{s^2+a^2} = s-a \tan^{-1}\left(\frac{s}{a}\right) \ .
\ee
Applying this to all the sub-integrals, the results are similar to those in 1D,
\bn
K_1 &=& \frac{1}{\omega+i\gamma} \left\{ 
-(\omega_g-\omega_{\nu})^{1/2}
\tan^{-1}\left(\frac{s}{\sqrt{\omega_g-\omega_{\nu}}}\right) \right. \nonumber \\
&+&  \left. (\omega_g-\zeta_{\nu})^{1/2}
\tan^{-1}\left(\frac{s}{\sqrt{\omega_g-\zeta_{\nu}}}\right) \right\} \ , \\
K_2 &=& \frac{1}{\omega+i\gamma} \left\{ 
-(\omega_g+\omega_{\nu})^{1/2}
\tan^{-1}\left(\frac{s}{\sqrt{\omega_g+\omega_{\nu}}}\right) \right. \nonumber \\
&+&  \left. (\omega_g+\zeta_{\nu})^{1/2}
\tan^{-1}\left(\frac{s}{\sqrt{\omega_g+\zeta_{\nu}}}\right) \right\} \ .
\en
Again, it will be useful to reverse the order in the radical in the first term in $K_1$,
assuming above the gap excitation.  Then we apply again,
$\sqrt{\omega_g-\omega_{\nu}}\rightarrow i \sqrt{\omega_{\nu}-\omega_g}$. Letting the
upper limit of integration $s_F\rightarrow \infty$, and inserting the limiting values
of the inverse tangents, there results
\bn
\label{Tnu1D-semi}
T_{\nu} &=&
\frac{\pi g_m}{2(\omega+i\gamma)^2}
\left\{ i \sqrt{\omega_{\nu}-\omega_g} - \sqrt{\omega_{\nu}+\omega_g} \right.  \nonumber \\
&& \left. + \sqrt{\omega_g-\zeta_{\nu}} + \sqrt{\omega_g+\zeta_{\nu}} \right\} .
\en
Note that for typical values of the parameters, both the real and imaginary parts of
$T_{\nu}(\omega)$ are positive.  This function derived has a peak in its real part, for
frequency near the gap frequency $\omega_g$.  Otherwise, it is very similar to the more complete
theory of (\ref{Tnux1D}) that includes the varying Fermi occupation factor.  That complete theory 
differs primarily in that the location of the peak in its real part is near the Fermi energy rather 
than the gap energy.


\begin{thebibliography}{99}
\bibitem{sign_convention} A positive Faraday rotation $\varphi$ corresponds to counterclockwise (CCW) rotation 
of the electric vector when the wave is approaching the observer (direction of rotation of the E-vector 
for left circular polarization).  A positive Faraday ellipticity ${\cal X}$ corresponds to CCW rotation  
of the E-vector in its elliptical motion in time, with minor axis $b$ over major axis $a$ satisfying 
${\cal X} = \tan^{-1}(b/a)$.  The E-vector rotates CCW with time, as would be the case for the 
left circular polarization component dominating.
\end{thebibliography}
\end{document}